\documentclass[preprints,article,accept,moreauthors,pdftex]{Definitions/mdpi} 
\firstpage{1} 
\pubvolume{9}
\issuenum{1}
\articlenumber{21}
\pubyear{2021}
\copyrightyear{2020}
\datereceived{} 
\dateaccepted{} 
\datepublished{} 
\hreflink{https://doi.org/} 
\usepackage{hyperref}
\hypersetup{colorlinks,allcolors=black}

\Title{Changing accretion geometry of Seyfert~1 Mrk~335 with {\it NuSTAR}: A comparative study}

\TitleCitation{Changing accretion geometry of Seyfert~1 Mrk~335 with {\it NuSTAR}: A comparative study}

\Author{Santanu Mondal*\orcidA{} and C. S. Stalin}

\AuthorCitation{Mondal, S.; Stalin, C. S.}

\address{%
$^{1}$ \quad Indian Institute of Astrophysics, II Block, Koramangala, Bangaluru, 560034, India; santanu.mondal@iiap.res.in}

\corres{Correspondence: santanu.mondal@iiap.res.in}

\abstract{We present a detailed spectral study of the narrow-line Seyfert~1 galaxy, Markarian~335, using eight epochs observations made between 2013 and 2020 with the {\it Nuclear Spectroscopic Telescope Array}. The source was variable during this period both in spectral flux and flow geometry. We estimated the height of the Compton cloud from the model fitted parameters for the whole observation period. This allowed us to investigate the underlying physical processes that drive the variability in X-rays. Our model fitted mass varies in a narrow range, between $(2.44\pm0.45-3.04\pm0.56)\times10^{7}M_\odot$, however, given the large error bars, is consistent with being constant and is in agreement with that known from optical reverberation mapping observations. The disk mass accretion rate reached maximum, 10\% of the Eddington rate during June 2013. Our study puts light on mass outflows from the system and also compares different aspects of accretion with X-ray binaries.}


\begin{document}

\section{Introduction}
Active galactic nuclei (AGN) host an accretion disk with an energetic X-ray producing Compton cloud so-called corona around a supermassive black hole at the center. The emitted X-ray radiation also acts as one of the most direct probes of accretion onto AGN. They show rapid flux and polarization variations which indicate that the observed high energy radiation mainly originates from the inner few to tens of Schwarzschild radii ($r_s=2GM_{\rm BH}/c^2$, $M_{\rm BH}$, G, and c are the mass of the black hole, gravitational constant and the speed of light respectively) from the event horizon of the black hole \citep{Mchr06}. The power-law (PL) component is widely accepted as the effect of inverse Compton (IC) scattering of thermally produced soft photons from the accretion disk by a corona of hot electrons at the inner edge of the disk \citep{Haar91,Zdzi00}. However, the geometry and size of the corona as well as the physical mechanisms governing the energy transfer between the disk and the corona are not well understood, which necessitates a physical model to better understand the observed signatures in them.

There are proposed models in the literature \citep{Esin97} which consider the corona to be the region between the truncation radius and the innermost stable circular orbit. However, it is not clear why the disk truncates at a certain radius from the central black hole and how the truncation radius is connected with the corona. Furthermore, most of the models in the literature \citep{Tita94,Garc13} use optical depth or the coronal temperature or the spectral index as a parameter to compute the spectrum from the corona. None of them are the basic physical quantities of accretion. According to two-component advective flow model, {\sc TCAF} \citep{Chak89,Chak95}, at a certain distance from the BH where gravitational force balances with the centrifugal force, shock forms by the low angular momentum, hot, sub-Keplerian halo, and satisfying Rankine-Hugoniot conditions, which is the region of the truncation of the disk. Apart from that as the two forces balances each other accreting matter virtually stops and its velocity goes down, which also increases the optical depth of the corona region. As the gravitational force dominates closer to the BH, velocity again increases and the flow passes through the inner sonic point and falls onto the BH, therefore the flow is transonic \citep{Chak89}. Beyond this shocked region, both the disk and halo matter pile up to decide the optical depth and temperature of the corona region. The same region also upscatters the incoming soft radiation from the disk via IC. A typical X-ray spectrum of AGN in the 2-10 keV shows primarily the signature of PL. Due to IC, the corona cools down and its area decreases, leading to low brightness level and a softer spectrum \citep{Mond13}. These physical properties and the dynamics of the flow makes the TCAF more favourable as a physical model than other existing models in the literature. In addition to the above, TCAF uses mass accretion rates and the mass of the BH as model parameters, which are the basic physical quantities of a flow. Furthermore, the same shock location which can explain the spectral features, can also be able to explain the temporal features from its oscillation \citep{Molt96,Chak15}. Therefore, among the various models available in the literature to understand the observed X-ray emission in AGN, we preferred to consider TCAF model for our study. 

TCAF model has mainly five parameters: (i) mass of the black hole ($M_{\rm BH}$), (ii) disk accretion rate ($\dot m_d$), (iii) halo accretion rate ($\dot m_h$), (iv) location of the shock ($X_s$), and (v) shock compression ratio (R). Increasing $\dot m_d$ keeping the other parameters fixed, increases the number of soft photons from the Keplerian disk \citep{Shak73} making the spectrum softer. Higher $\dot m_h$ increases the corona temperature and hardens the spectrum. Similarly increase of either $X_s$ or R lead to hardening of the spectrum. However, in reality all parameters change in multidimensional space. Therefore such variations may not sustain due to non-linear change in optical depth and coronal temperature in different epochs. A detailed parameter study has been made in Ref. \cite{Chak97}. In recent years, TCAF model is implemented in XSPEC and has been used widely to study both X-ray binaries (XRBs) \citep{Debn14,Mond14,Debn15} and AGN \citep{Mand08,Nand19,Mond21}. However, there are physical processes which are not yet incorporated in the current TCAF model e.g. jet, bulk motion Comptonization and the spin effect of the BH.

Apart from the above components, many AGN exhibit extra features in their X-ray spectra that likely originate from the accretion disk \citep{Fabi10}. The so-called X-ray reflection features produced by the accretion disk might be because of its illumination from the central compact object, such as a magnetically dominated corona residing above the surface of the disk \citep[e.g.][]{Gale79,Haar91,Merlo01}. A prominent feature in the X-ray spectrum is the Fe K$\alpha$ emission line at around 6.4 keV which is due to fluorescence in a dense and relatively cold medium \citep{Barr85,Nand89,Poun90}. The shape of the Fe K$\alpha$ line is a powerful probe of the general relativistic effects close to the BH \citep{Fabi89,Laor91}, leading to estimates of the spin of the BH \citep{Bren06,Reyn08,Reyn13}. Indeed, observations over the last decades by X-ray satellites have uncovered relativistically broadened Fe K$\alpha$ line in the spectra of several AGN \citep{Iwas96,Ball03,Risa13} and black hole XRBs \citep[e.g.][]{Mond16}. Models of X-ray reflection spectra \citep{Ross05,Garc13} show additional features that collectively can constrain the ionization state, heating, and cooling of the illuminated surface \citep{Ross99}.

Mrk~335 is a bright narrow-line Seyfert 1 (NLS1) galaxy at a redshift z = 0.026 \citep{Huch99}, with a black hole mass, $M_{\rm BH}=2.6\times10^7 M_\odot$ \citep{Grie12} and showing ionized absorption in the X-ray and ultra-violet (UV) bands \citep{Long13}. It has been observed to go into extremely low-flux states \citep{Grup07} where the soft X-ray flux dropped by a factor of up to $\sim 30$, while the hard
flux dropped only by a factor of $\sim 2$ in 7 years. Using observations from XMM-Newton, Ref. \cite{Grup08} found signatures of absorption and high reflection fraction in the spectrum of Mrk~335 in its lowest flux state, whereas at the intermediate flux state the observed X-ray spectra is explained well by blurred reflection model without the requirement of variable absorption \citep{Gall13}. Ref. \cite{Kara13} estimated a high-frequency lag of $\sim 150$ s between the continuum dominated energy bands and the iron line and soft excess components for this source. This time lag suggests that the continuum source is located very close to the central black hole and that relativistic effects are supposed to be worth considering. 
Using combined {\it Swift} and {\it NuSTAR} observations, Ref. \cite{Wilk15b} interpreted the X-ray flare during 2014 as arising from the vertical collimation and ejection of the X-ray emitting corona at a mildly relativistic velocity, that causes the continuum emission to be relativistically beamed away from the disk. Ref. \cite{Wilk15a} discussed the effect of the geometry of the corona on the relativistically blurred X-ray reflection arising from the accretion disk, which can also explain the variability in between low and high-flux states. 
Ref. \cite{Gall18} performed the structure function analysis using long term {\it Swift} optical/UV and X-ray data of Mrk~335 and showed that the X-ray low flux state could be due to the physical changes in the corona or absorption, whereas the variability in the optical/UV band is more consistent with the thermal and dynamic timescales associated with the accretion disk. 
Ref. \cite{Gall19}, studied the low-flux state data using {\it XMM-Newton} and observed absorption lines from a highly ionized outflowing wind. 

Several studies on Mrk 335 clearly show that the system is highly variable leaving a range of possibilities in its dynamical behavior and emission features. Mrk 335 is known to show change in geometry \citep{Wilk15a}, such a change in geometry can be found from a disk-based model fit such as TCAF to data. It is not clear to date if there is any change in the mass accretion of the source, and if so, what effects it can have on the geometry of the corona in the source? To investigate the above, in this work we chose this object and carried out an analysis of the X-ray data on the source acquired by {\it NuSTAR} during the period 2013-2020. This paper is organized as follows: in the next section, we describe the observation details and the analysis procedures. In \S 3, we explain our results obtained from model fits to the data and summarize our conclusions in the final section. 
\section{Observation and Data analysis}
In the present manuscript, we analyzed archival data\footnote{https://heasarc.gsfc.nasa.gov/docs/cgro/db-perl/W3Browse/w3browse.pl} of {\it NuSTAR} observations of Mrk 335 made during 2013-2020. During this time interval Mrk 335 was observed nine times, out of which in one observation, data quality is not good, so we considered the remaining eight observations. The details of the observations are listed in 
\autoref{table:observation}. The {\it NuSTAR} data in the energy range of 3.0$-$30~keV (as the data is noisy above 30 keV), were extracted 
using the standard {\sc NUSTARDAS v1.3.1} \footnote{https://heasarc.gsfc.nasa.gov/docs/nustar/analysis/} software. We 
ran {\sc nupipeline} task to produce cleaned event lists and 
{\sc nuproducts} for generation of the spectra. We used a region of 
$80^{\prime\prime}$  for the source and $100^{\prime\prime}$ for the background 
using ds9. The data were grouped by {\it grppha} command, with a 
minimum of 30 counts per energy bin. The same binning was used for all the 
observations. For spectral analysis of the data we used {\sc XSPEC}\footnote{https://heasarc.gsfc.nasa.gov/xanadu/xspec/} \citep{Arna96} version 12.8.1. 
We used the absorption model 
TBABS \citep{Wilm00} with the Galactic hydrogen column density 
fixed at 3.6$\times$~10$^{20}$~atoms~cm$^{-2}$ \citep{Kalb05} throughout the analysis. 

\begin{table}
\centering
\caption{\label{table:observation} Log of observations.}
\begin{tabular}{cccc}
\hline
	Date    &MJD    & OBSID         &Exposure (s) \\
\hline
2013-06-13 	&56456  &60001041002	&21299\\
2013-06-13 	&56456&60001041003	&21525 \\  
2013-06-25 	&56468&60001041005	&93028\\
2014-09-20 	&56920&80001020002	&68908 \\
2018-07-10 	&58309&80201001002	&82257\\
2020-06-06 	&59006&90602619004	&30156\\
2020-06-07 	&59007&90602619006	&30495\\
2020-06-08 	&59008&90602619008	&22452\\
\hline
\end{tabular}
\end{table}

\section{Results and Discussion}
\begin{figure}
    \centering{
    \includegraphics[height=8truecm,angle=0]{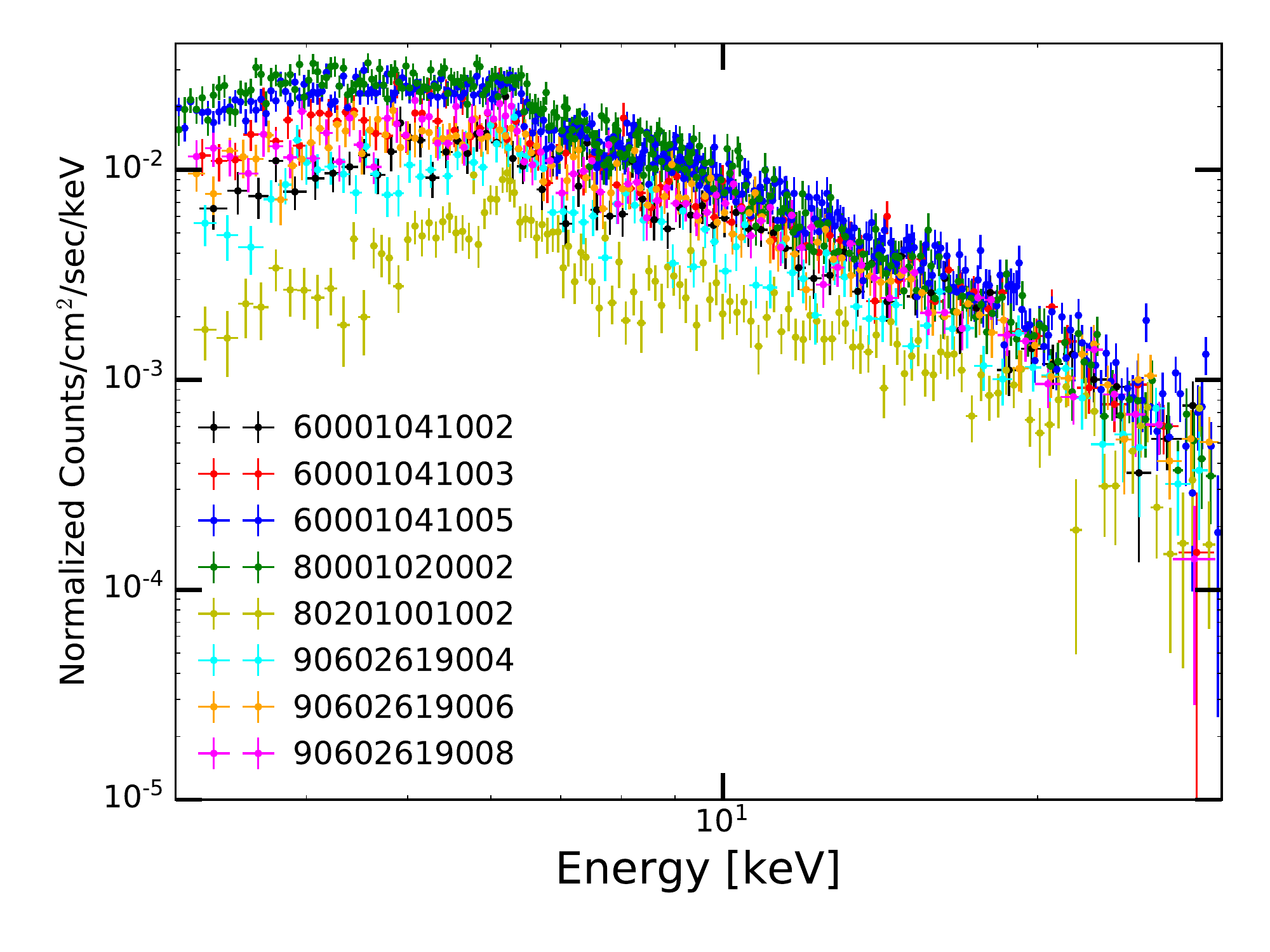}}
    \caption{Spectra of all the eight observations in the 3$-$30 keV energy range. Variability of the source is evident. \label{fig:Combined8Spec}}
\end{figure}

We fit the {\it NuSTAR} observations of Mrk~335 taken between 2013 to 2020. During this observation period the source showed significant variability in spectral flux (see \autoref{fig:Combined8Spec}). We did spectral analysis using different models. First, we performed spectral fitting using a simple powerlaw (PL) model as {\sc tbabs(gauss+pl)}, where the PL index ($\Gamma$) varies from 0.72-1.84, a change by a factor greater than two, which is clearly an indication of significant change in the flow behavior and corona properties. The model fitted parameters are provided in \autoref{table:powerlaw}. Our model fits show that the line energy and width vary in a broad range, which point to complex geometry changes and gravity effects (more especially line broadening) in the source. 
In addition to the simple PL model, we also modeled the observed spectra using the complex phenomenological model {\sc pexrav}  \citep{Magd95}  and the physical model TCAF \citep{Chak95}. The results of PEXRAV model fits are given in \autoref{table:pexrav}. Our PL model fits indicate that the source passed through hard and soft spectral states similar to XRBs, where the timescale for changing from one state to other is in the order of few days to weeks, which is the viscous timescale \citep[][and references therein]{Mond17}. We found a similar timescale when the corona changed from its maximum size to minimum, which is around 12 days. However, as the viscous timescale for AGN is in the order of few hundred to few Myr (scaled by the mass of the BH), it may not be possible to observe the similar timescale for other AGNs. In future we aim to explore this aspect of accretion to estimate the viscosity of the flow.

\begin{figure*}
    \centering{
    \includegraphics[height=8truecm,angle=270]{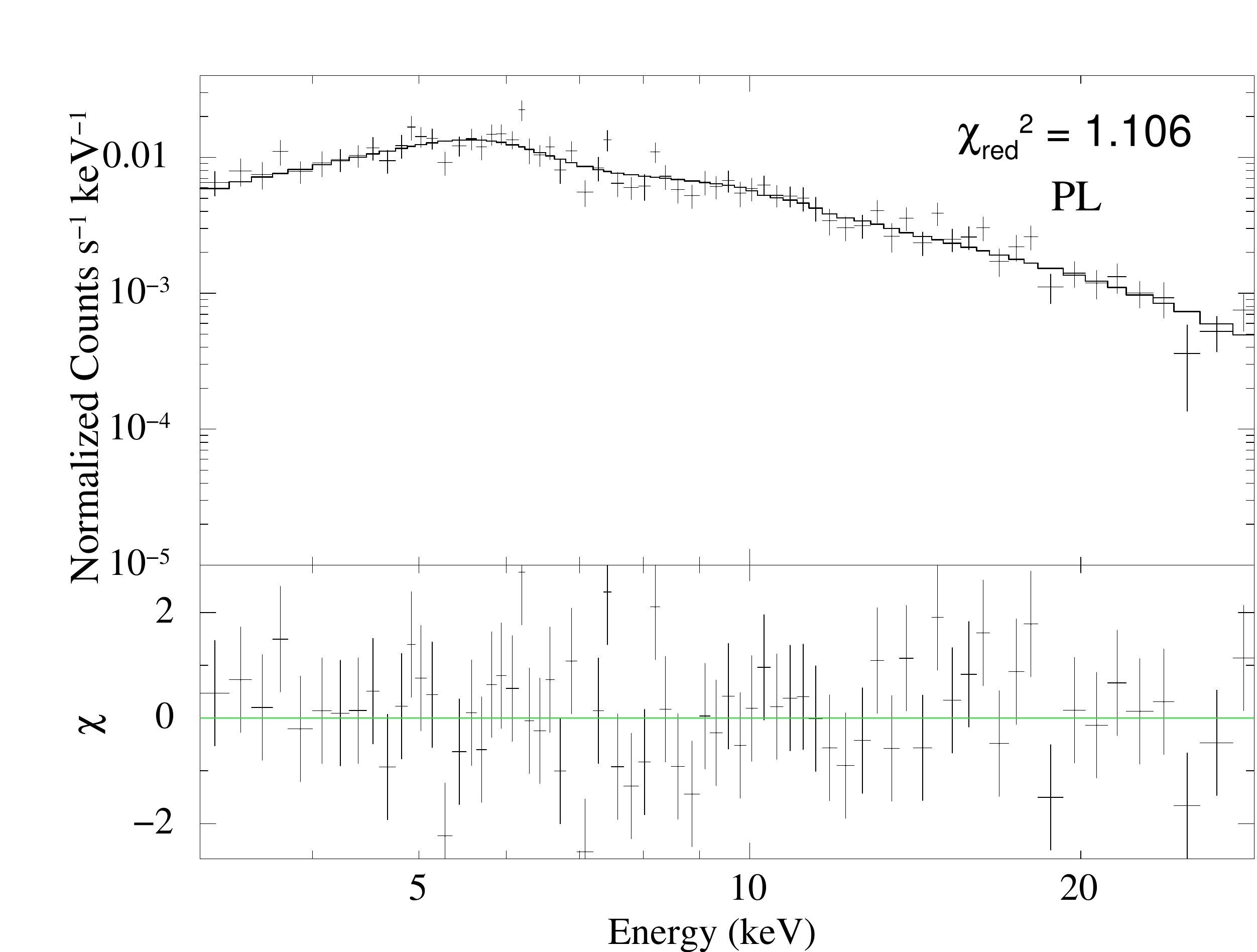}
    \includegraphics[height=8truecm,angle=270]{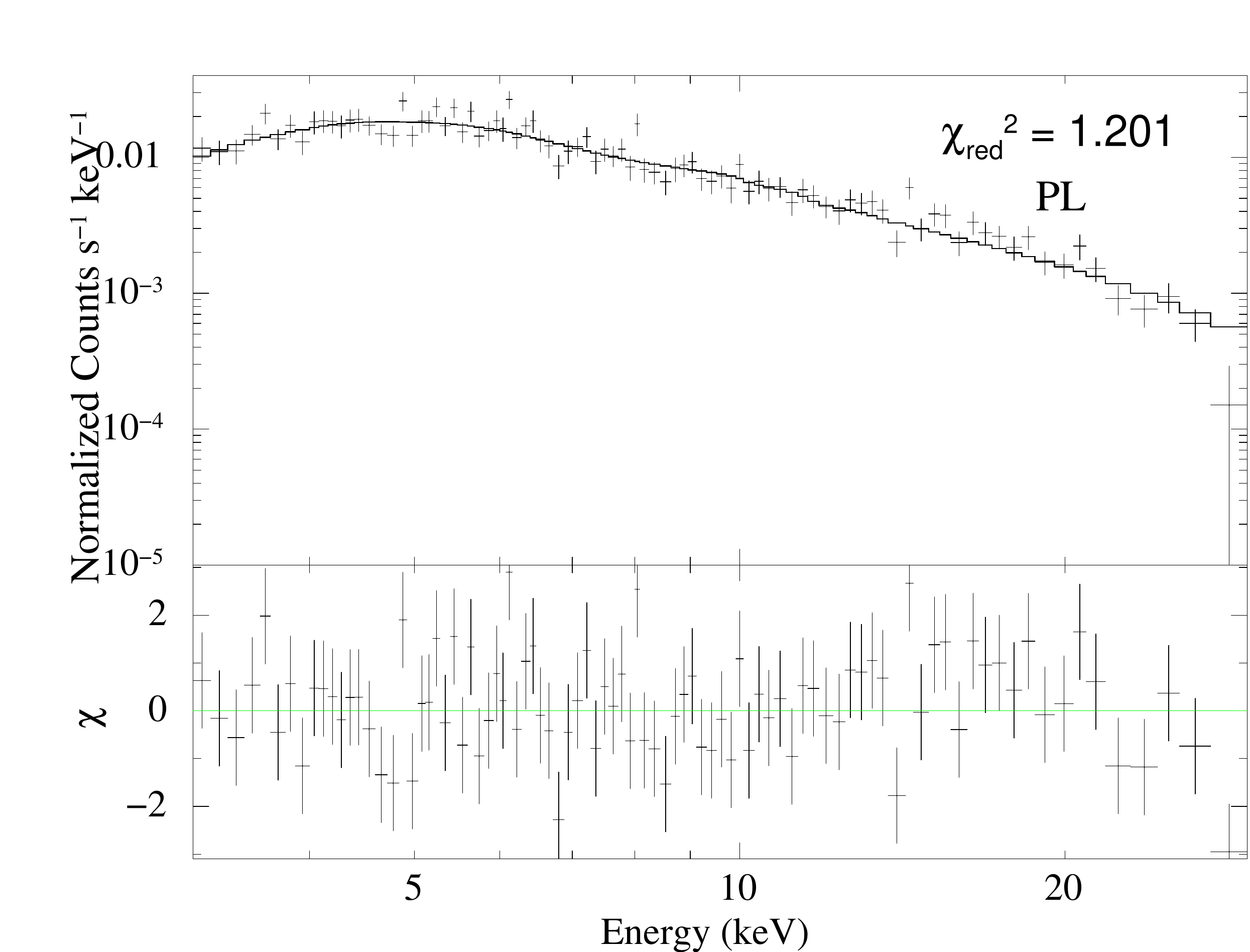}}
   \centering{
   \includegraphics[height=8truecm,angle=270]{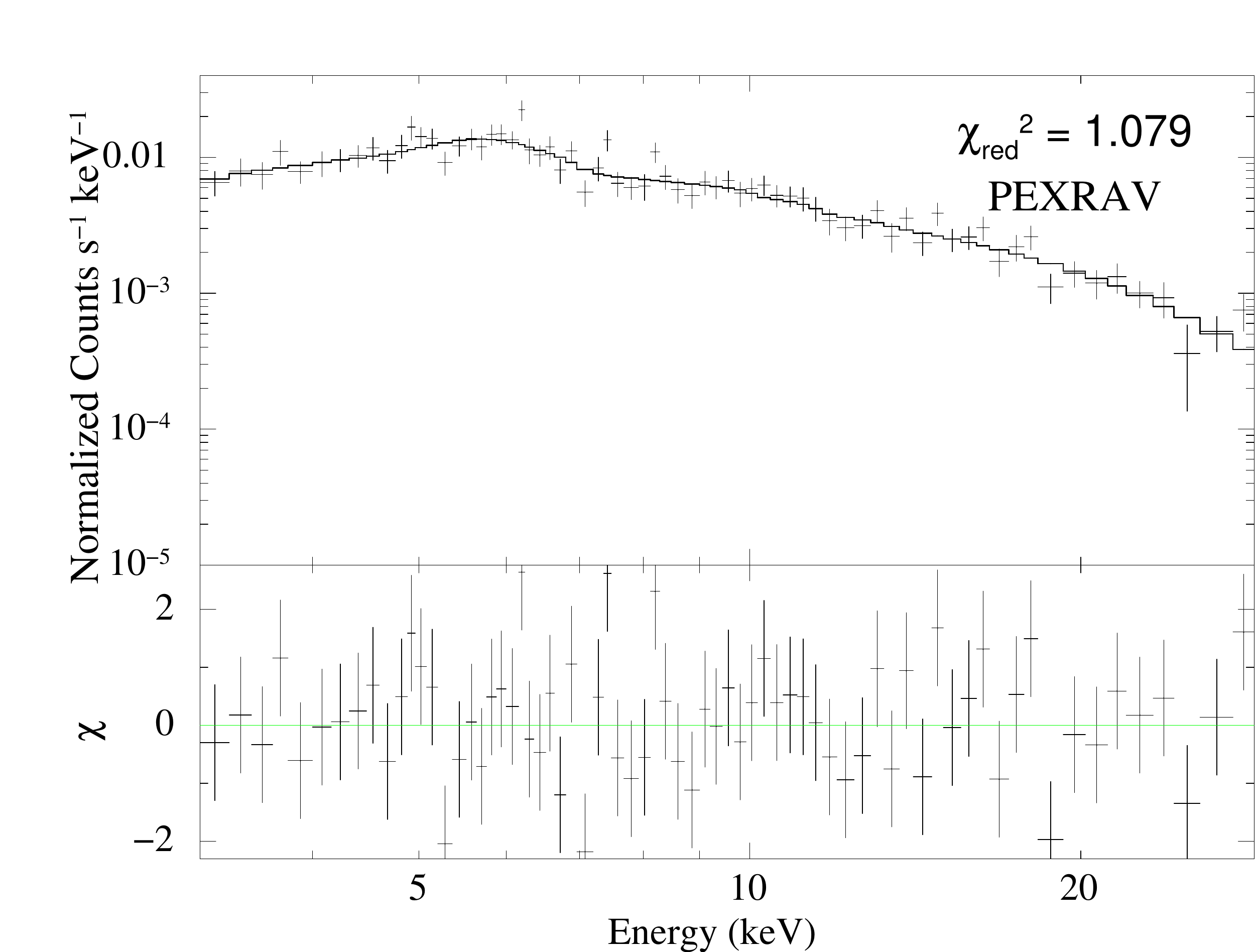}
   \includegraphics[height=8truecm,angle=270]{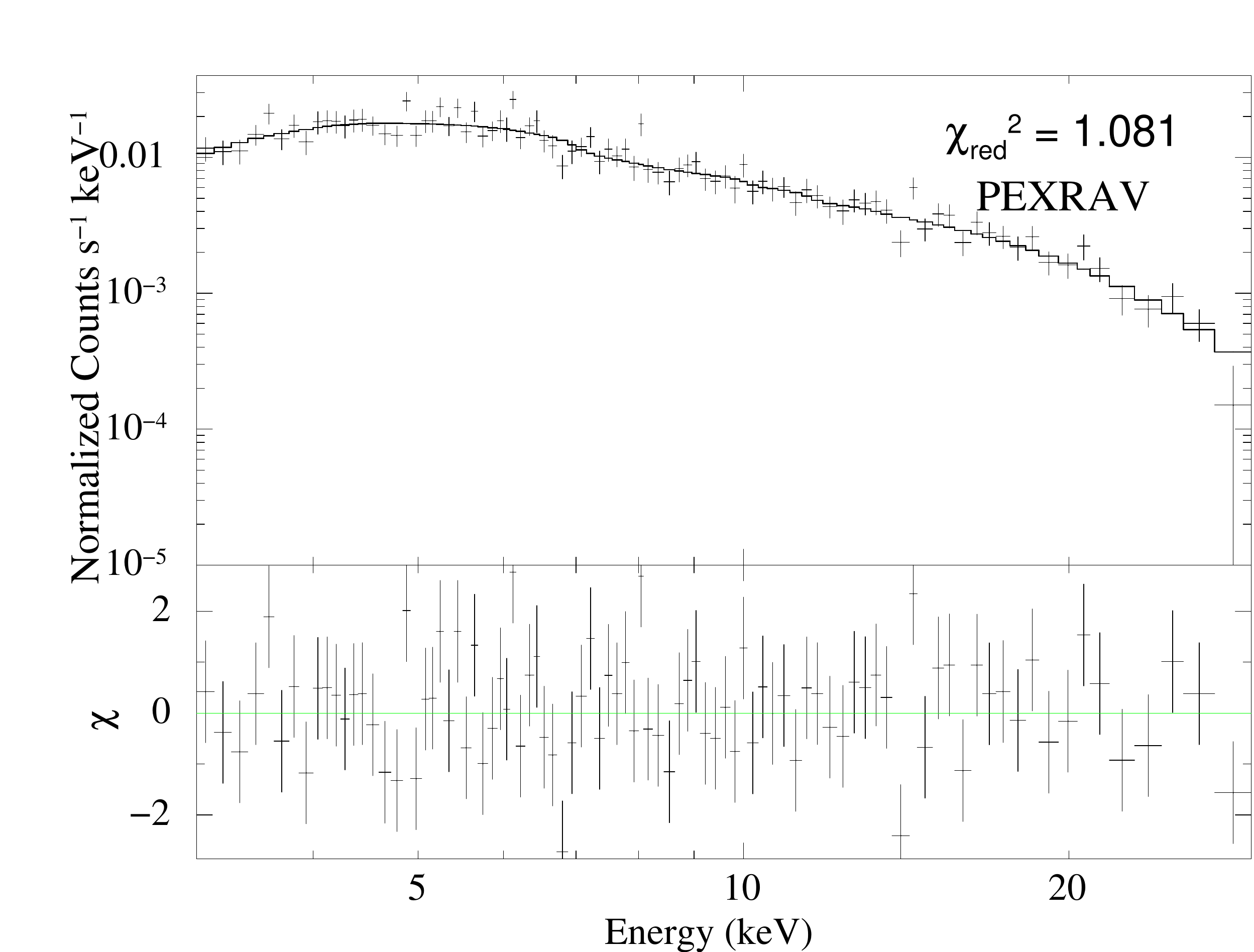}}
   \centering{
   \includegraphics[height=8truecm,angle=270]{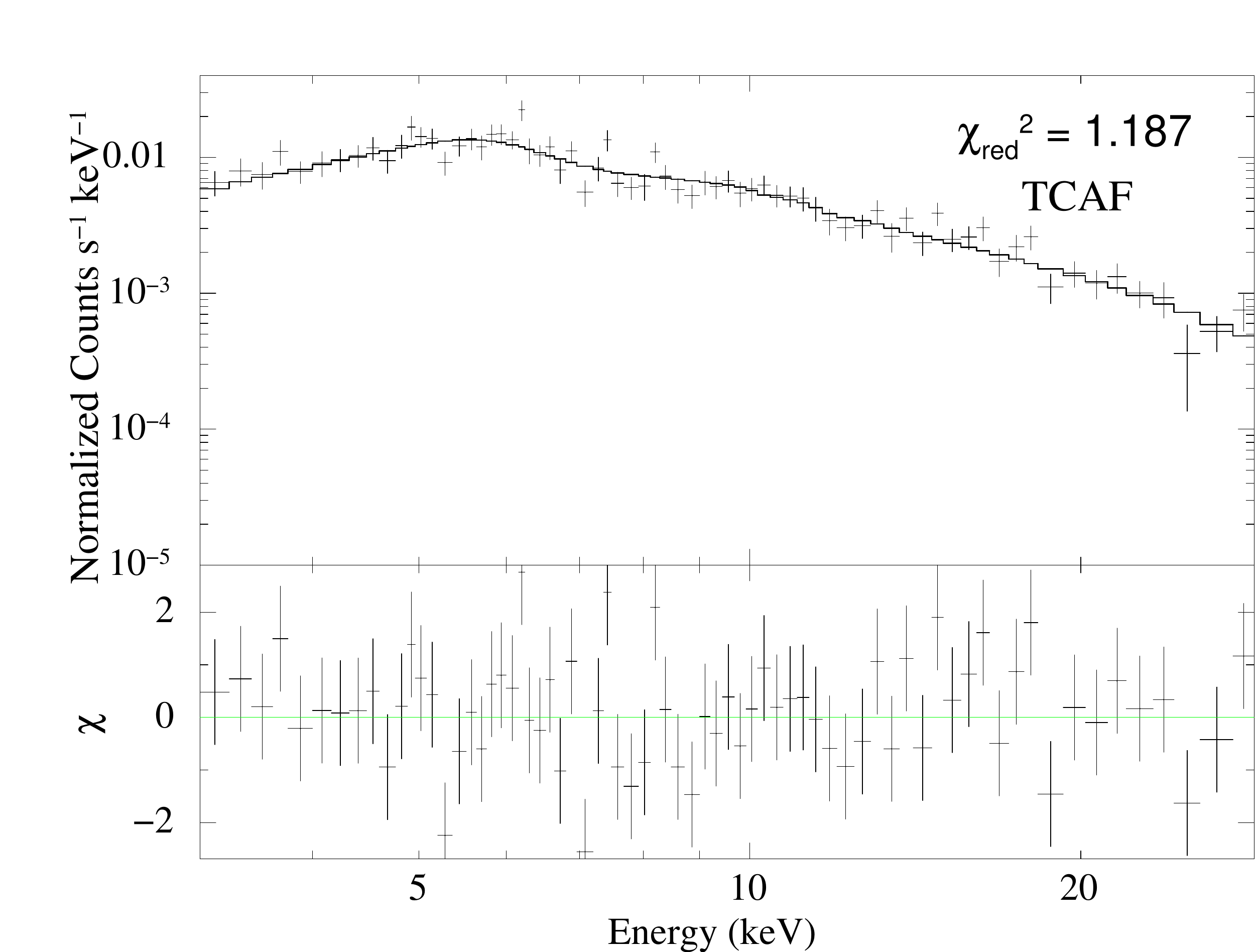}
   \includegraphics[height=8truecm,angle=270]{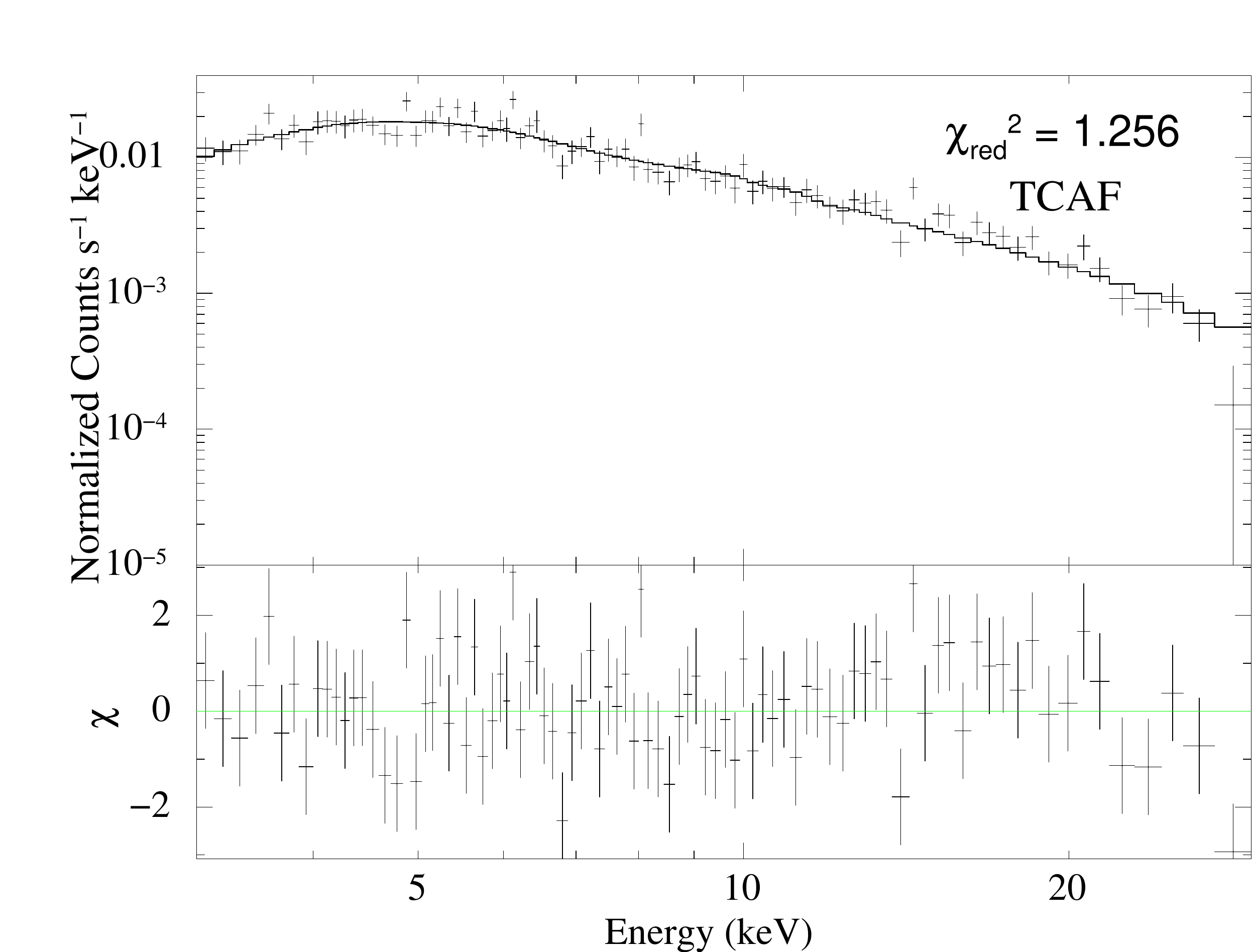}}
    \caption{Sample fit of 3$-$30 keV spectra with {\sc PL}, {\sc pexrav}, and {\sc tcaf}  models for OBSIDs 60001041002 (left panel) and 60001041003 (right panel) along with the residuals. \label{fig:specPLPEXRAVTCAF}}
\end{figure*}

\begin{figure*}
    \centering{
    \includegraphics[height=8truecm,angle=270]{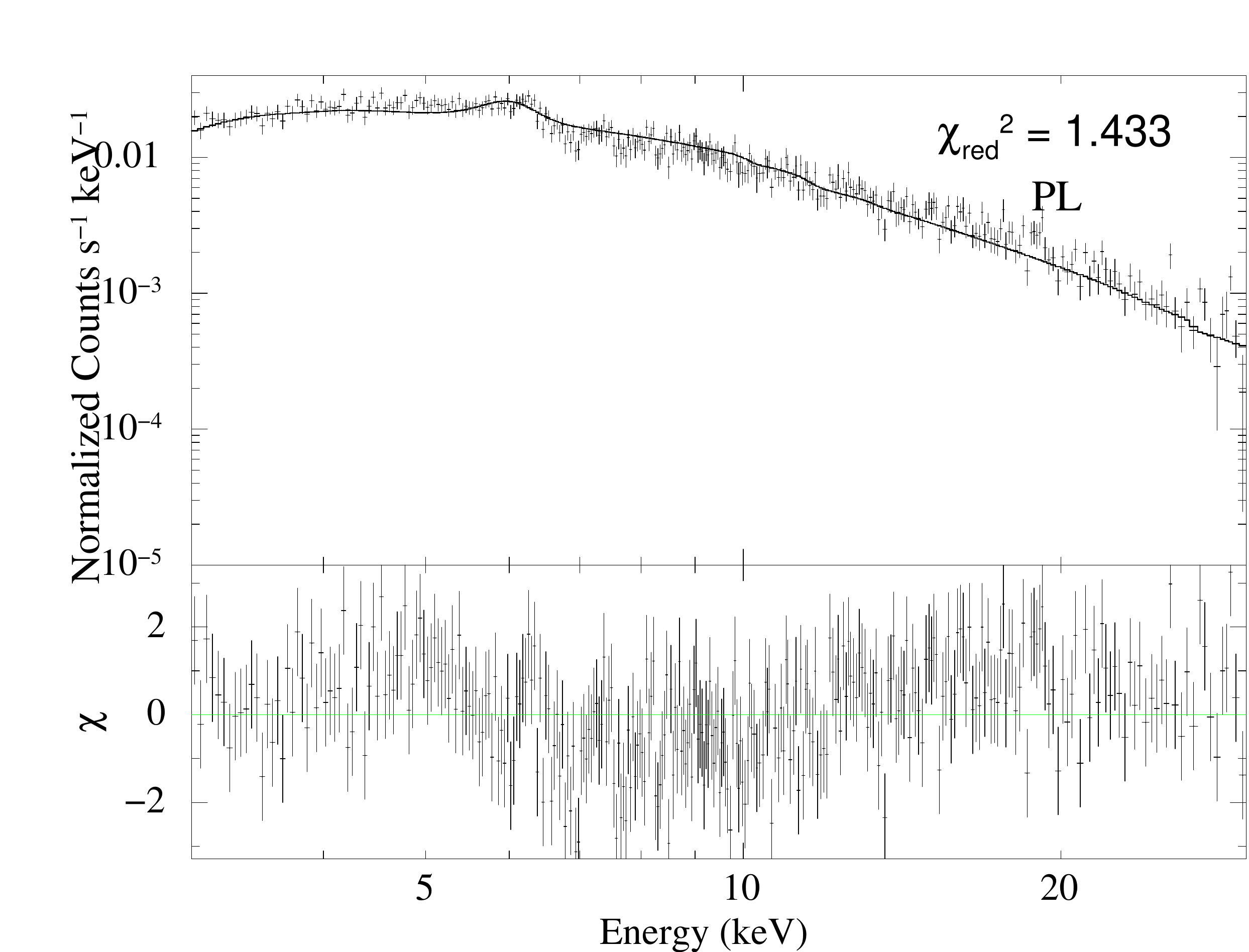}
    \includegraphics[height=8truecm,angle=270]{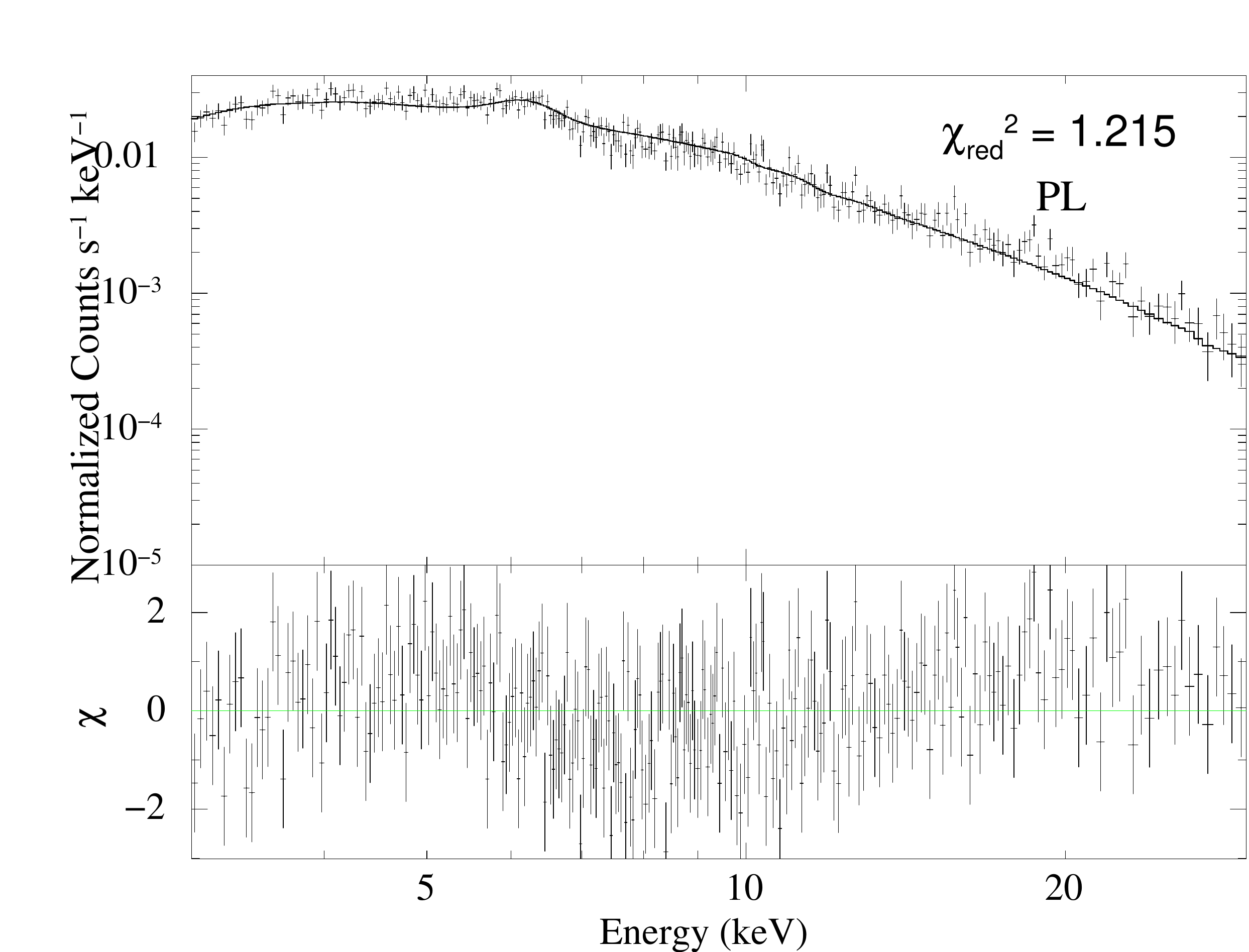}}
   \centering{
   \includegraphics[height=8truecm,angle=270]{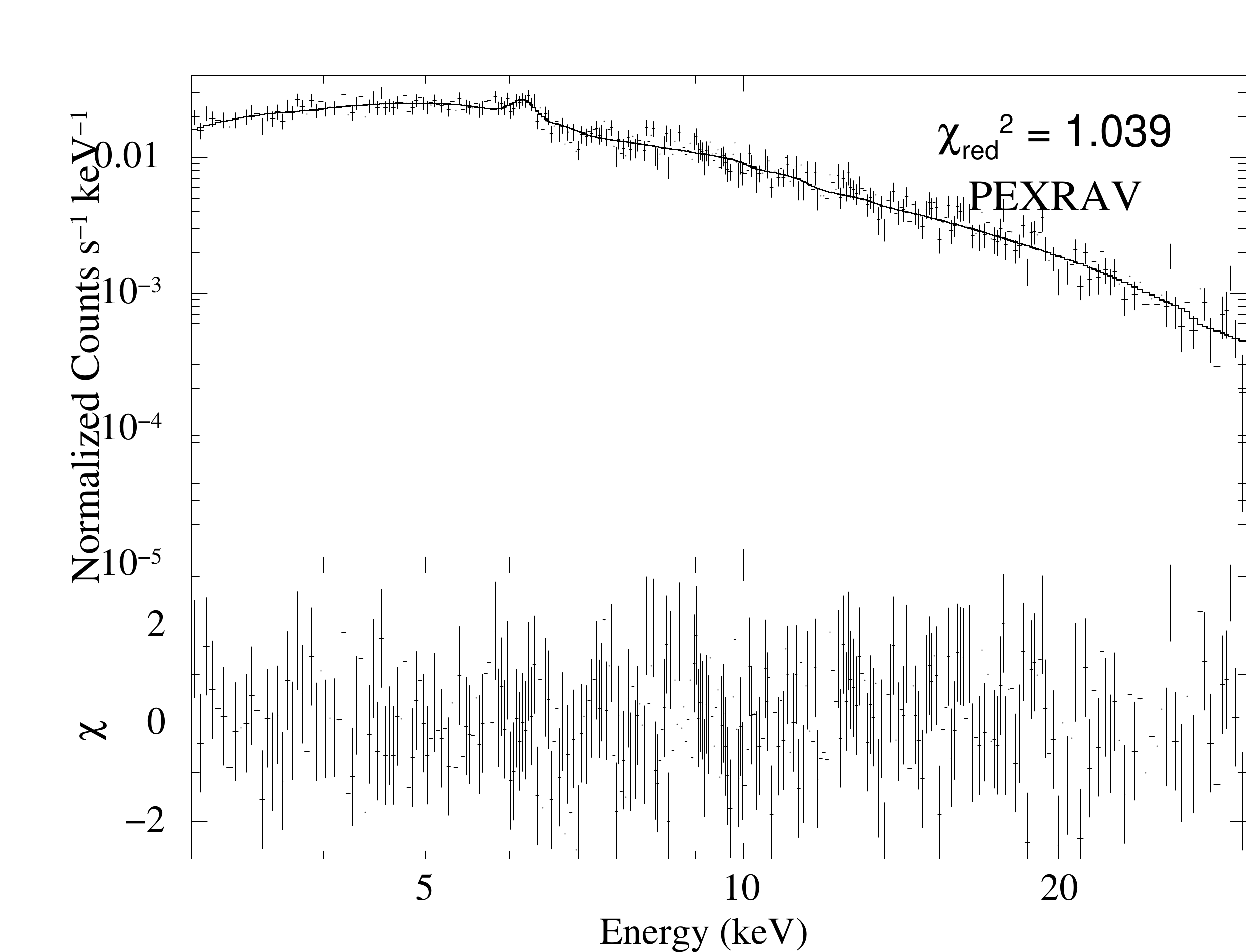}
   \includegraphics[height=8truecm,angle=270]{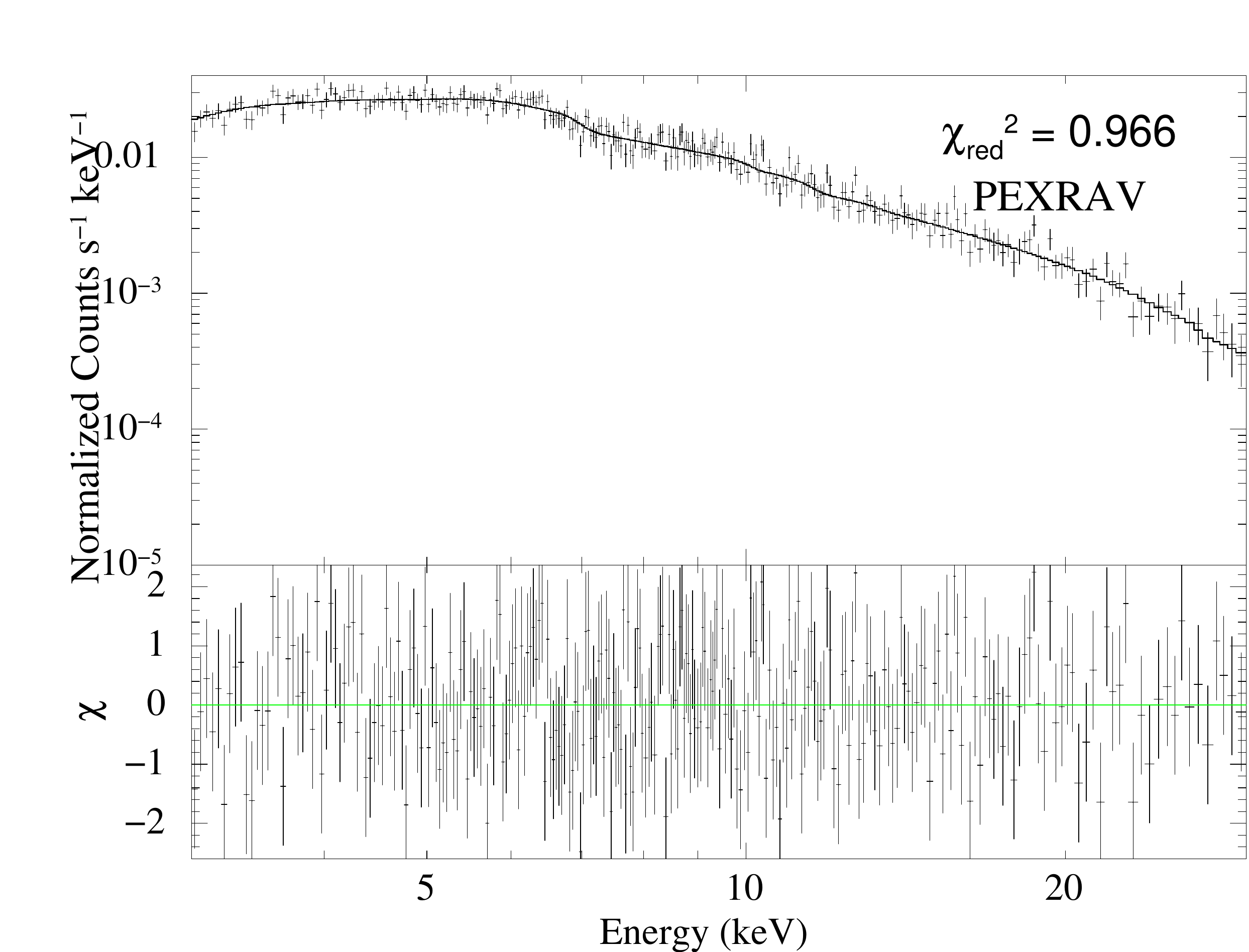}}
   \centering{
   \includegraphics[height=8truecm,angle=270]{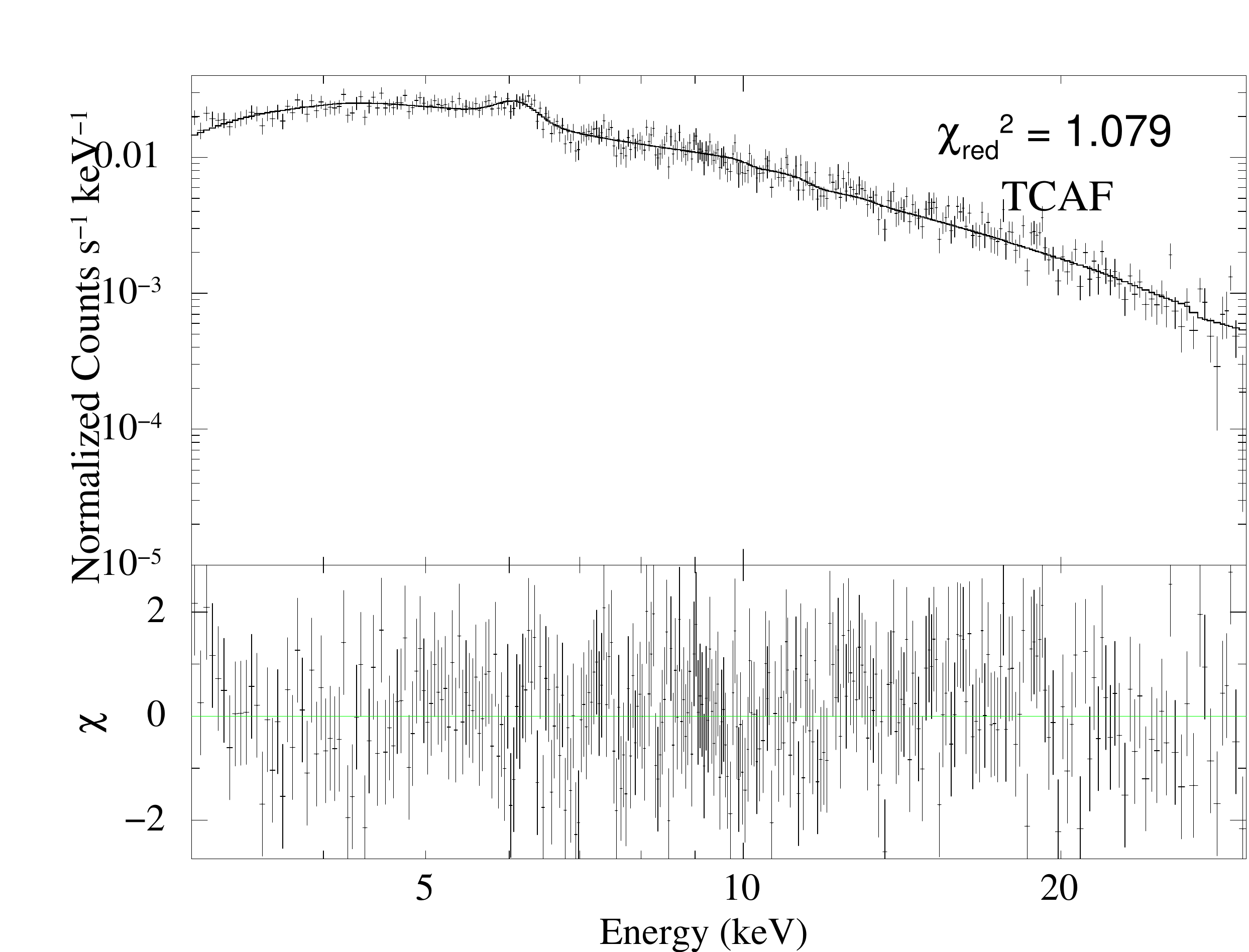}
   \includegraphics[height=8truecm,angle=270]{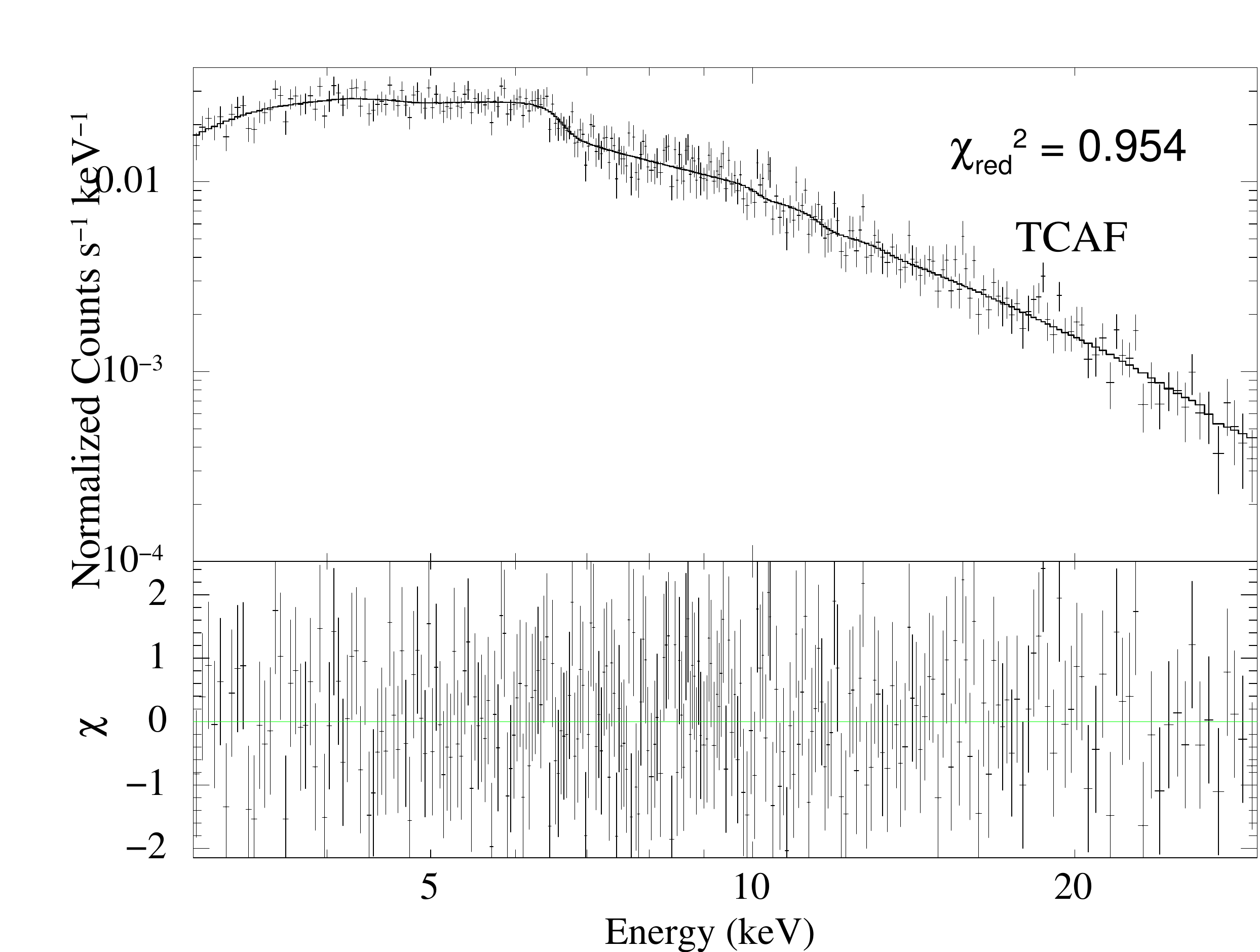}}
    \caption{Same as \autoref{fig:specPLPEXRAVTCAF} but for the OBSIDs 60001041005 (left panel) and 80001020002 (right panel) along with the residuals. \label{fig:specPLPEXRAVTCAF-2}}
\end{figure*}

From PEXRAV model fits, we found the reflection fraction to be high with $R_{\rm ref} > 0.9$ for all the epochs. When the Fe abundance ($A_{\rm Fe}$) was made as a free parameter, $A_{\rm Fe}$, was found to vary from sub-Solar to super Solar values during the epochs analysed here, which is quite unphysical. It has been pointed out by Ref. \cite{Mast20}, reflection model fits to observed X-ray spectra are generally found to yield high $A_{\rm Fe}$ values. Also, according to Ref. \cite{Garc15}, the high $A_{\rm Fe}$ obtained from model fits to data could be due to some unknown physical effects being overlooked in current models. Therefore, for PEXRAV model fits, we froze $A_{\rm Fe}$ to solar value. For the observation day MJD=56468, we required one extra Gaussian line component at low energy $\sim 5$ keV. From the line component and its width it can be seen, that both narrow and broad lines are present in this epoch. We kept the disk inclination ($i_{\rm inc}$) fixed to $45^\circ$ throughout \citep{Chai15}. Next we applied the physical model {\sc tcaf} and having the form {\sc tbabs(tcaf+gauss)} to extract the accretion flow parameters. For all observations goodness of the fit is more or less similar to {\sc pexrav} model. The {\sc tcaf} model fitted parameters are provided in \autoref{table:tcaf}. For some observations, better fits were achieved after adding some additive model components which are detailed in the table. For majority of the epochs, the $\chi^2/dof$ obtained from PEXRAV and TCAF model fits are in agreement, which for three epochs, the $\chi^2/dof$ from TCAF fits are marginally larger than those obtained from PEXRAV model fits. This might be due to differences in the spectral components between models e.g. TCAF does not include line features, that may give relatively higher $\chi^2/dof$. In \autoref{fig:specPLPEXRAVTCAF}, we show different model fitted spectra for the OBSID 60001041002/3 in the left/right panels of the figure. \autoref{fig:specPLPEXRAVTCAF-2} shows the same spectral fitting but for the OBSID 60001041005 (left panel) and 80001020002 (right panel). The rest of the observational fits are shown in Appendix.

\begin{table*}
\centering
\caption{\label{table:powerlaw} Best-fitting parameters with the {\sc tbabs(gauss+pl)} model. }
\begin{tabular}{|l|l|l|l|l|l|ccccc|}
\hline
OBSID&$\Gamma_{\rm PL}$ & $N_{\rm PL} (\times10^{-4})$ &$E_{g}$ &$\sigma_{g}$ &$\chi^2/dof$ \\
      &  &$ph./keV/cm^2/s $& [keV] & [keV]   & \\
\hline
60001041002 &$1.09\pm0.08$&$ 2.15\pm0.46$&$5.47\pm0.18$&$0.79\pm0.21$&71.91/65\\ 
60001041003 &$1.12\pm0.10$&$ 2.79\pm0.76$&$4.21\pm0.45$&$1.52\pm0.34$&102.1/85\\
60001041005 &$1.66\pm0.02$&$13.58\pm0.54$&$6.00\pm0.99$&$0.28\pm0.06$&452.94/316\\
80001020002 &$1.84\pm0.02$&$20.92\pm0.91$&$6.20\pm0.13$&$0.34\pm0.08$&320.71/264\\
80201001002 &$0.72\pm0.06$&$ 0.40\pm0.06$&$6.07\pm0.07$&$0.60\pm0.08$&136.14/124\\
90602619004 &$1.24\pm0.06$&$ 2.50\pm0.33$&$6.08\pm0.07$&$0.41\pm0.08$&107.16/79\\
90602619006 &$1.17\pm0.09$&$ 2.93\pm0.71$&$4.83\pm0.31$&$1.26\pm0.30$&82.39/106\\
90602619008 &$1.46\pm0.05$&$ 6.07\pm0.61$&$6.10\pm0.06$&$0.13\pm0.09$&92.29/80\\ 
\hline
\end{tabular} 

\noindent{$\Gamma_{\rm PL}$ and $N_{\rm PL}$ are the {\sc pl} index and normalization. $E_g$ and $\sigma_g$ are the Gaussian line energy and width respectively.}
\end{table*}

\begin{table*}
\small
\centering
\caption{\label{table:pexrav} Best-fitting parameters with the {\sc tbabs(pexrav+gauss)} model.}
\begin{tabular}{|l|l|l|l|l|l|l|cccccc|}
\hline
OBSID&$\Gamma$ & $E_{\rm cut}$ &  $R_{\rm ref}$ & $E_{g}$ &$\sigma_{g}$ &$\chi^2/dof$ \\
      & & [keV] &   &  [keV] & [keV]   & \\
\hline
60001041002&$1.44\pm0.03$&$31.55\pm6.51$&$3.40\pm1.17$&$5.66\pm0.09$&$0.61\pm0.21$&68.01/63\\
60001041003&$1.12\pm0.02$&$16.56\pm2.42$&$4.85\pm1.07$&$3.40\pm0.17$&$1.99\pm0.37$&89.72/83\\
$60001041005^{a}$&$1.95\pm0.10$&$96.30\pm20.87$&$2.40\pm0.41$&$6.20\pm0.06$&$0.007\pm0.003$&323.19/311\\
80001020002&$2.11\pm0.09$&$107.90\pm30.15$&$2.33\pm0.92$&$5.50\pm0.17$&$0.84\pm0.17$&253.39/262\\
80201001002&$0.87\pm0.03$&$20.40\pm3.93$&$3.68\pm0.81$&$6.09\pm0.08$&$0.62\pm0.08$&128.70/122\\
90602619004&$1.68\pm0.03$&$54.63\pm11.71$&$3.47\pm1.12$&$6.11\pm0.08$&$0.39\pm0.09$&96.97/77\\
90602619006&$1.39\pm0.02$&$69.19\pm15.89$&$0.92\pm0.17$&$5.20\pm0.28$&$1.01\pm0.36$ &81.39/104\\
90602619008&$1.99\pm0.04$&$32.69\pm4.01$&$7.67\pm1.07$&$6.12\pm0.06$&$0.013\pm0.003$&78.82/78\\
\hline
\end{tabular} 

\noindent{$\Gamma$, $E_{\rm cut}$, and $R_{\rm ref}$ are model PL photon index, cutoff energy, and reflection scaling factor respectively. $E_g$ and $\sigma_g$ are the Gaussian line energy and width. We kept the disk inclination fixed at 45$^\circ$ \citep{Chai15} and $A_{\rm Fe}$ at Solar abundance value.}
\noindent{a: Gaussian component was added at $5.02\pm0.18$~keV with line width $0.64\pm0.20$~keV. The origin of this line is not clear yet, however, a similar line was observed for Mrk~335 earlier \citep{Ezhi20}}.
\end{table*}

\begin{table*}
\small
\centering
\caption{\label{table:tcaf} Best-fitting parameters with the {\sc tbabs(tcaf+gauss)}.}
\begin{tabular}{|l|l|l|l|l|l|l|l|l|cccccccc|}
\hline
OBSID&$M_{\rm BH}\times10^{7}$&$\dot m_d$ &$\dot m_h$ & $X_s$ &  R & $E_g$ & $\sigma_g$  &$\chi^2/dof$ \\
      &[$M_\odot$]&$[\dot M_{\rm Edd}$] &$[\dot M_{\rm Edd}$] & [$r_s$] &   & [keV] & [keV]  & \\
\hline
60001041002&$2.94\pm0.63$&$0.0024\pm0.0008$&$2.000\pm0.725$&$22.43\pm9.96 $&$2.45\pm0.54$&$5.58\pm0.19$&$0.69\pm0.21$&72.45/61\\
60001041003&$3.02\pm0.60$&$0.0026\pm0.0010$&$1.158\pm0.331$&$43.72\pm12.97$&$5.78\pm1.23$&$4.18\pm0.61$&$1.53\pm0.52$&101.75/81\\
$60001041005^{a}$&$2.44\pm0.45$&$0.1001\pm0.0433$&$1.781\pm0.143$&$6.24\pm0.94  $&$5.21\pm1.26$&$6.10\pm0.05$&$0.22\pm0.07$&333.54/309\\
$80001020002^{b}$&$2.93\pm0.65$&$0.0059\pm0.0001$&$3.117\pm0.181$&$8.66\pm1.66  $&$1.49\pm0.17$&$2.13\pm0.52$&$2.27\pm0.95$&244.18/256\\
$80201001002^{c}$&$2.79\pm0.70$&$0.0145\pm0.0011$&$3.017\pm0.144$&$33.53\pm7.62 $&$4.24\pm1.39$&$6.02\pm0.13$&$0.66\pm0.15$&130.29/117\\
$90602619004^{d}$&$3.02\pm0.98$&$0.0059\pm0.0014$&$2.945\pm0.365$&$21.25\pm3.19 $&$2.79\pm0.62$&$6.01\pm0.08$&$0.51\pm0.09$&87.65/72\\
90602619006&$3.04\pm0.56$&$0.0024\pm0.0007$&$1.978\pm0.150$&$21.54\pm2.57 $&$2.20\pm0.26$&$4.78\pm0.62$&$1.29\pm0.49$&82.35/102\\
90602619008&$2.87\pm0.95$&$0.0045\pm0.0012$&$1.540\pm0.259$&$22.22\pm2.49 $&$1.20\pm0.13$&$5.53\pm0.19$&$0.70\pm0.22$&95.42/76\\
\hline
\end{tabular} 
\noindent{$\dot{m_h}$, and $\dot{m_d}$ represent {\sc tbabs(tcaf+Gauss)} model fitted sub-Keplerian (halo) and Keplerian (disk) rates respectively. $X_s$, and $R$ are the model fitted shock location and shock compression ratio values respectively. $E_g$ and $\sigma_g$ are the Gaussian line energy and width.}
\noindent{a: Gaussian component was added at $3.44\pm0.07$~keV with line width $1.55\pm0.21$~keV.}
\noindent{b: {\sc diskline} \citep{Fabi89} component was added with line energy at $6.12\pm0.10$~keV, powerlaw of emissivity ($\beta$)=$-2.98\pm0.70$, inner radius ($R_{\rm in}$)=$34.6\pm24.4$ $r_g (GM/c^2)$, and the disk inclination was frozen at $45^\circ$.}
\noindent{c: {\sc zxipcf} \citep{Reev08} multiplicative model component is used for parameters: column density $(N_{H})=68\pm10.91 \times 10^{22} cm^{-2}$, covering fraction $(f_c)=0.45\pm0.05$, $log\xi=1.2\pm0.3$.}
\noindent{d: Gaussian component was added at $4.16\pm0.11$~keV with line width $0.29\pm0.14$~keV.}
\end{table*}

We used {\sc tcaf} model fitted parameters to estimate the geometrical height of the shock ($H_{\rm shk}$). The height of the corona is basically the height of the shock, as the shock is the boundary layer of the corona. We estimated ($H_{\rm shk}$) using the equation below \citep[see also][]{Debn14}:
\begin{equation}
    H_{\rm shk}=\left[\frac{\gamma (R-1) X_s^2}{R^2}\right]^{1/2},
\end{equation}
where, $\gamma$ is the adiabatic index, $X_s$ is the location of the shock or the boundary of the corona, and R is the shock compression ratio ($=\rho_+/\rho_-$), a ratio of the density of the downstream to the density of the upstream of the flow. 

\begin{figure} 
    \centering{
    \includegraphics[height=10truecm]{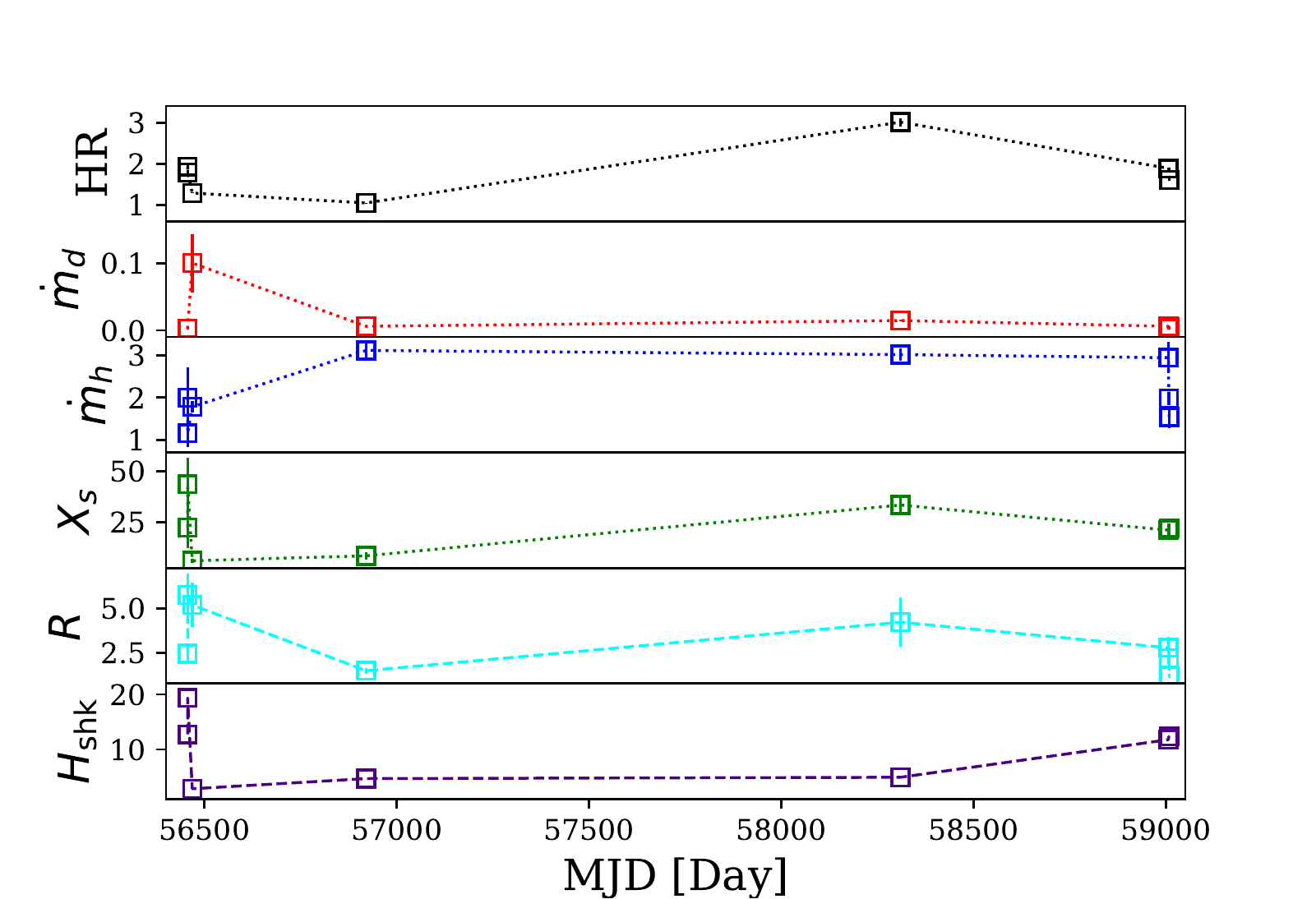}}
    \caption{TCAF model fitted parameters evolution with observation days are shown along with HR estimated from the PL model fitted fluxes.}
    \label{fig:AllParsHR}
\end{figure}

In \autoref{fig:AllParsHR}, we show the variation of various model parameters obtained from TCAF model fits as well as the variation of hardness ratio (HR) with time. The halo rate was always higher than the disk rate by an order of magnitude or more which implies that the flow was dominated by the low angular momentum matter, which might be accreted from the wind or host galaxies. The $\dot m_d$ was found to vary by a factor of $\sim$40 from 0.24\% to  10\%  of the Eddington rate and the $\dot m_h$ varied between 1.2-3.1~Eddington rate. The size of the shock location ($X_s$) or corona changed significantly and reached up to 6~$r_s$ during 25 June 2013 (60001041005) and 8$r_s$ during 2014 (80001020002). For the OBSID 60001041005, we required an extra broad (1.55 keV) Gaussian component at low energy 3.44~keV, which was also found to be broadened by strong gravitational effects \citep{Laor91}. For the OBSID 80001020002, we required an additive {\sc diskline} \citep{Fabi89} component at line energy 6.12 keV with emissivity powerlaw index ($\beta$) $\sim -3.0$ and $R_{\rm in}=24.4 r_g (GM/c^2)$ as the current TCAF model does not include line emission due to gravitational effect in the fit. It is worth noting that the $X_s$ is achieved for the highest $\dot m_d$, which might be due to the cooling effect inside the corona. As the accretion rate increased more disk photons got intercepted by the corona. This cooled the corona, causing the shock to move inwards and thus causing a significant change in the size of the corona. Such change in the corona could also be due to the activity of the jet, which extracted a significant amount of thermal energy and contracted the corona \citep{Chak99}. Therefore it is also be possible to establish a jet-disk connection using TCAF model fits to AGN spectra similar to XRBs \citep[][and references therein]{Chak02,Mond14b,Chat19}. 

Such movement of the shock can give rise to the observed variability in spectral and temporal properties \citep{Chak93}. During 2014, $\dot m_h$ further increased and made the corona hotter and shock the receded slightly. During 2018, shock further moved away, the corona has become bigger in size, therefore intercepts more soft photons, generates high energy photons, which can also ionize the disk atmosphere. For this particular observation, we required partial covering fraction model ({\sc zxipcf}), with a covering factor ($f_c$=0.45) and low ionization parameter ($log\xi$=1.2). For the OBSID 90602619004, during 2020, we required another narrow line (0.29 keV) component, when the shock again moved inward. The model fitted shock compression ratio also varied in a broad range, which also dictates the strength of the shock and the optical depth of the corona. The 5$^{th}$ panel of \autoref{fig:AllParsHR} (from top) shows the variation of this parameter. The bottom panel of \autoref{fig:AllParsHR}, shows the estimated height of the corona, which changes significantly for the whole observation period.

Mass of the BH is one of the parameters in TCAF. Therefore, keeping it as a free parameter during TCAF model fits to the observed spectra will yield $M_{\rm BH}$ value for Mrk 335. We kept $M_{\rm BH}$ as a free parameter during our fitting and we found $M_{\rm BH}$ to vary in range between ($2.44\pm0.45$ to $3.04\pm0.56$) $\times 10^7 M_\odot$. Considering the large error bars the derived BH mass is consistent to be constant. Our derived $M_{\rm BH}$ is consistent with that estimated by \citep{Grie12}  using optical reverberation mapping observations, however, an order of magnitude larger than that obtained by \citep{Mast20}. Such a low mass found by \citep{Mast20} could be due to them not considering an expanded corona in their model fit to the data. 
Our results on $M_{\rm BH}$ also shows that by fitting accretion disk based models such as TCAF to observed X-ray spectra  one can estimate mass of BH in AGN. Also, we found that freezing $M_{\rm BH}$ to some fixed value from the above range and redoing the fit has negligible effect on other derived parameters, such as $\dot m_d$, $\dot m_h$, $X_s$, and R.
 
The viscous time scale in AGN is much longer compared to that of XRBs. At small time scales (in the order of few days or month) significant change in the accretion rate is not expected, similar is the case with the heating and cooling rates. Thus at such timescales, variability is expected to be less. This is in fact observed in the radio-quiet category of AGN and accretion disk based models can explain the observed flux variations in them. However, this might not be the case in the radio loud category of AGN, as relativistic jets play a dominant role in the flux variations observed in them. Even in those sources as the ejection of jets extract thermal energy from the corona, the change in the shock location can be drastic in them though the cooling is less. Results on this will be reported elsewhere. 

\section{Conclusion}
In the present paper, we analyzed {\it NuSTAR} data of Mrk~335, observed between 2013-2020 to understand the dynamics of the flow and the accretion behavior of the source. We found that the source was significantly variable both in spectral flux, accretion geometry, and reflection fraction. We also found the geometry of the corona to change between epochs. Such a change can affect the reflection fraction and the spectral energy distribution of the illuminating radiation, and consequently the ionization rate.
From the TCAF model fits we obtained the dynamics of the flow along with the geometry change of the corona and accretion rates. 
In our TCAF model fits, the line profiles due to relativistic effects were taken into account using an additive
DISKLINE model where ever needed rather than using throughout for all observations. Among other models, even though the XILLVER model has relativistic and reflection effects incorporated, the accretion dynamics and the origin of corona are still lacking in the model. Therefore we did not take into account such models in our current study. Below we provide the key findings on the variable nature of the source:

\begin{enumerate}

\item During 2013, the corona at the inner region has changed significantly from its elongated stage ($\sim r_s$) to destruction stage $\sim 6 r_s$, consistent with Ref. \cite{Wilk15b}. As the compression ratio has changed and the corona has contracted significantly, it can be possible that jet/mass outflow was launched around June 13, 2013, therefore a significant amount of thermal energy has been extracted by the jet/outflow and the corona contracted. 

\item The observation during 2014, required a blurred reflection component along with {\sc tcaf} to fit the spectra. This is quite natural as the corona contracted, the inner edge of the disk moved significantly inward, therefore the gravitational effect became dominant \citep{Fabi89,Laor91} and blurred the Fe~K$\alpha$ line. We also required a broad Gaussian line component at $\sim 2$keV. 

\item The HR was roughly constant during 2013, and is also similar to that obtained during 2014. 

\item There is a significant change in $\Gamma_{\rm PL}$ between 2013 and 2014 spectra, which can be due to sudden change in accretion rates. During this period disk accretion had increased by a factor of a few and also the size of the corona contracted significantly.

\item The steepening of the emissivity profile of Mrk~335 indicates that the corona is compact for this source \citep{Wilk15b}. In our study, we found that the size of the corona is indeed small and compact during 25$^{th}$ June 2013 and 20$^{th}$ September 2014 (see \autoref{table:tcaf}). It is also noticeable that the height of the corona reduced significantly during these two epochs.

\item During 2014, spectral flux between 3-30 keV changed/increased by a factor of $\sim 3$ compared to 2013 and 2018. This could be due to an increase in accretion rates as well as the change in corona. As the accretion rate increased (in 25$^{th}$ June 2013 and 10$^{th}$ July 2018), the number of soft photons increased, thereby increasing the cooling rate, i.e reduction of more energy from the corona by the seed photons from the Keplerian disk \citep[see theoretical aspects in][]{Mond13}. It should be noted that though the shock location changed significantly, the other parameters triggered that change, mainly the increase in disk accretion rate, therefore the cooling rate. This also infers that not only the shock location but other parameters are equally important to explain the observed variability.

\item During 2018 and onward, the corona and $H_{\rm shk}$ again elongated. During this period the HR also increased. This also implies that there is a correlation between HR and the geometry of the corona.  

\item During 2018, to take into account the disk ionization effects along with TCAF, we required partial covering, {\sc zxipcf} model to better fit the data. The model fit showed an absorption column density with $N_H=6.8\times 10^{23}$ $\text{cm}^{-2}$ is present. The fit also required a low ionization with, $log\xi=1.2$~erg cm $\text{s}^{-1}$ with partial covering fraction of 0.45. This added component also indicates the presence of mass outflow from the system, which is evident in the monitoring observations of the source in optical/UV and X-rays \citep[][and references therein]{Park19,Komo20}.

\item The reflection fraction, R$_{\rm ref}$, is measured as the ratio of the photon fluxes from the blurred reflection and powerlaw continuum model components. From {\sc pexrav} model fitting we found this value $> 0.9$. 

\item The mass of the black hole which is kept as a free parameter is found to vary in a very narrow range (2.44-3.04)$\times 10^7 M_\odot$, and considering the error bars is consistent with a constant. This is in agreement  with that of Ref. \cite{Grie12}.

\item Earlier studies on observed X-ray variability inferred the origin of flux variations to changes in the primary powerlaw continuum possibly exhibited through intrinsic variations in the corona, or possible changes in its size or location \citep{Gall13,Wilk15b,Sarm15}. This is in agreement with our present findings.

\end{enumerate}

\vspace{6pt} 

\authorcontributions{``Conceptualization, methodology, formal analysis, writing---original draft preparation, Santanu Mondal; conceptualization, supervision, writing---review and editing, C. S. Stalin."}

\funding{Not Applicable}

\institutionalreview{Not Applicable}

\informedconsent{Not Applicable}

\dataavailability{We used archival data for our analysis in this manuscript. Appropriate links are given in the manuscript. For the details of the data fitting, one can directly contact to the first author.} 

\acknowledgments{We thank both referees for their constructive and insightful suggestions that improved the quality of the manuscript. SM acknowledges Ramanujan Fellowship (\# RJF/2020/000113) by SERB, Govt. of India. This research has made use of the {\it NuSTAR} Data Analysis Software (NUSTARDAS) jointly developed by the ASI Science Data Center (ASDC), Italy and the California Institute of Technology (Caltech), USA. This research has also made use of data obtained through the High Energy Astrophysics Science Archive Research Center Online Service, provided by NASA/Goddard Space Flight Center.}

\conflictsofinterest{``The authors declare no conflict of interest.'' } 

\end{paracol}
\reftitle{References}

\appendixstart
\appendix
\section{}
Here we show the spectral fitting plots for the rest of the observations using all three models.

\begin{figure*}[t]
    \centering{
    \includegraphics[height=8truecm,angle=270]{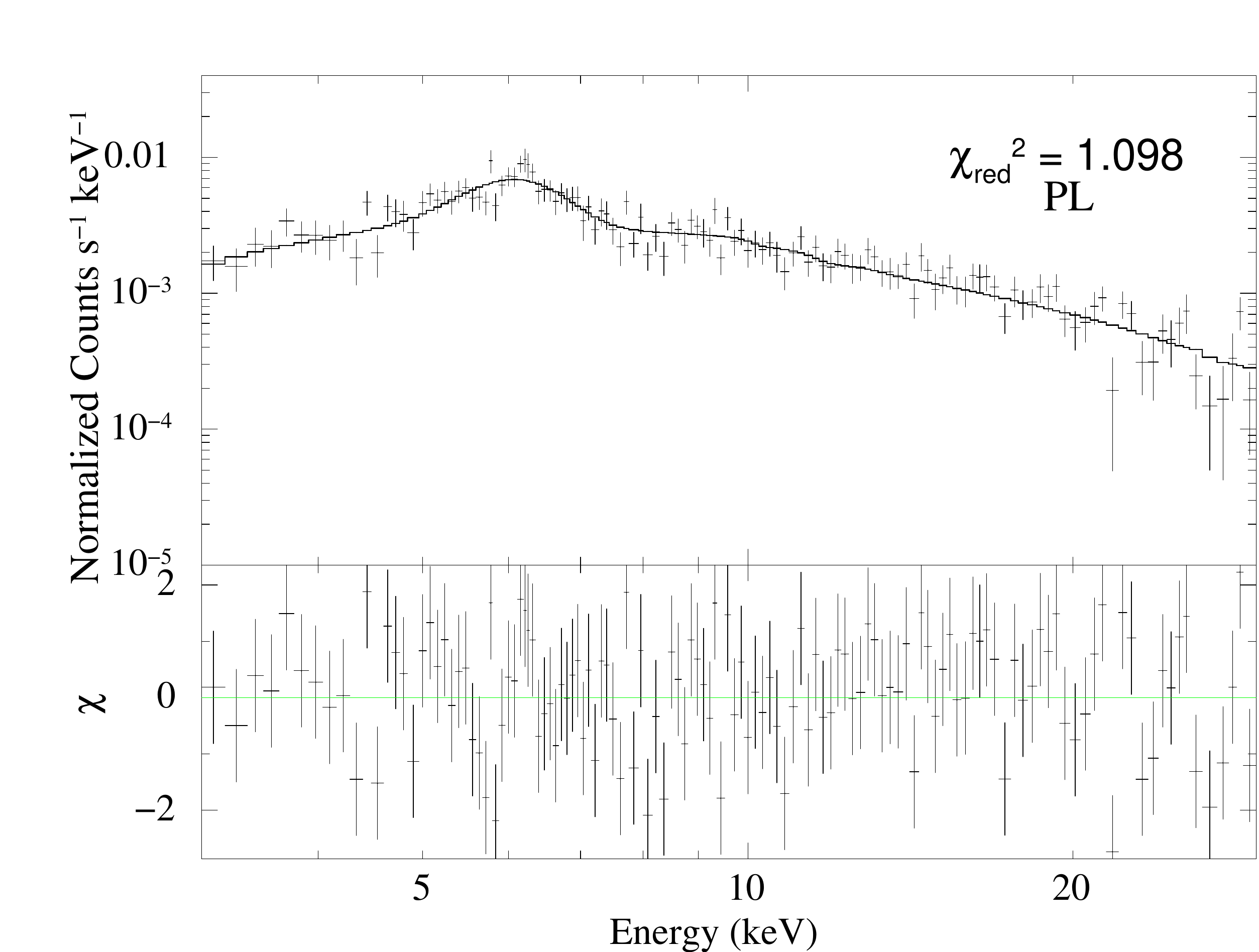}
    \includegraphics[height=8truecm,angle=270]{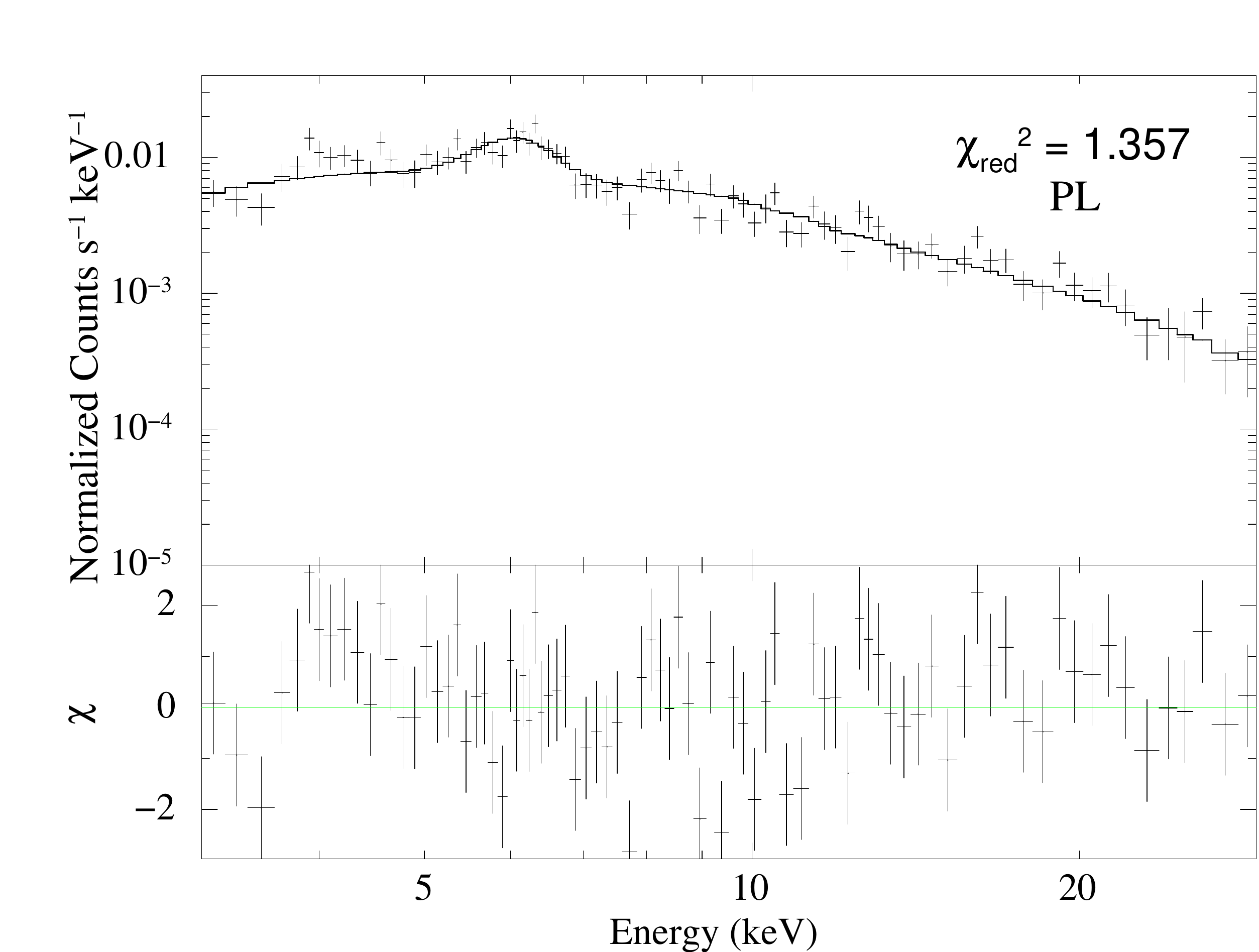}}
   \centering{
   \includegraphics[height=8truecm,angle=270]{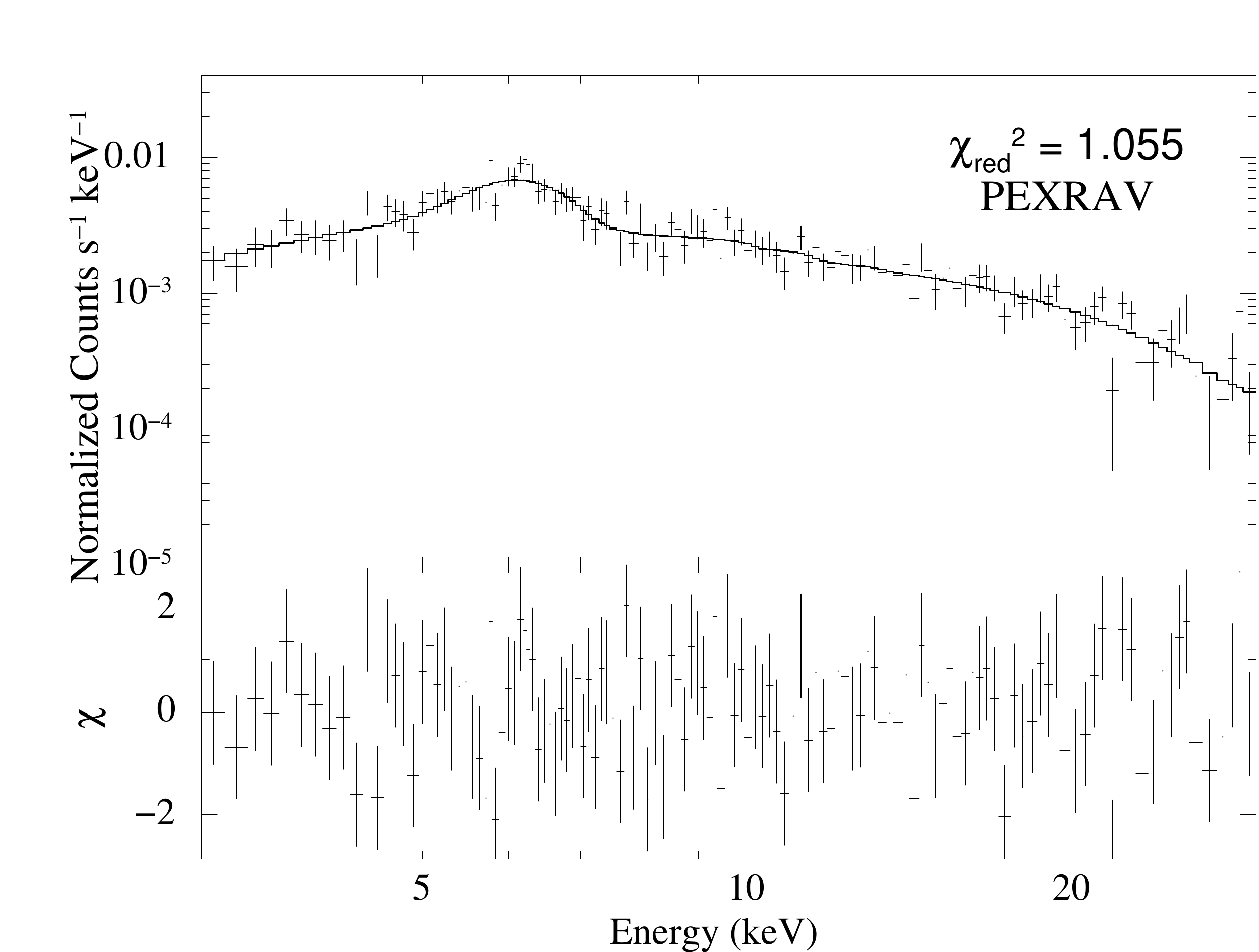}
   \includegraphics[height=8truecm,angle=270]{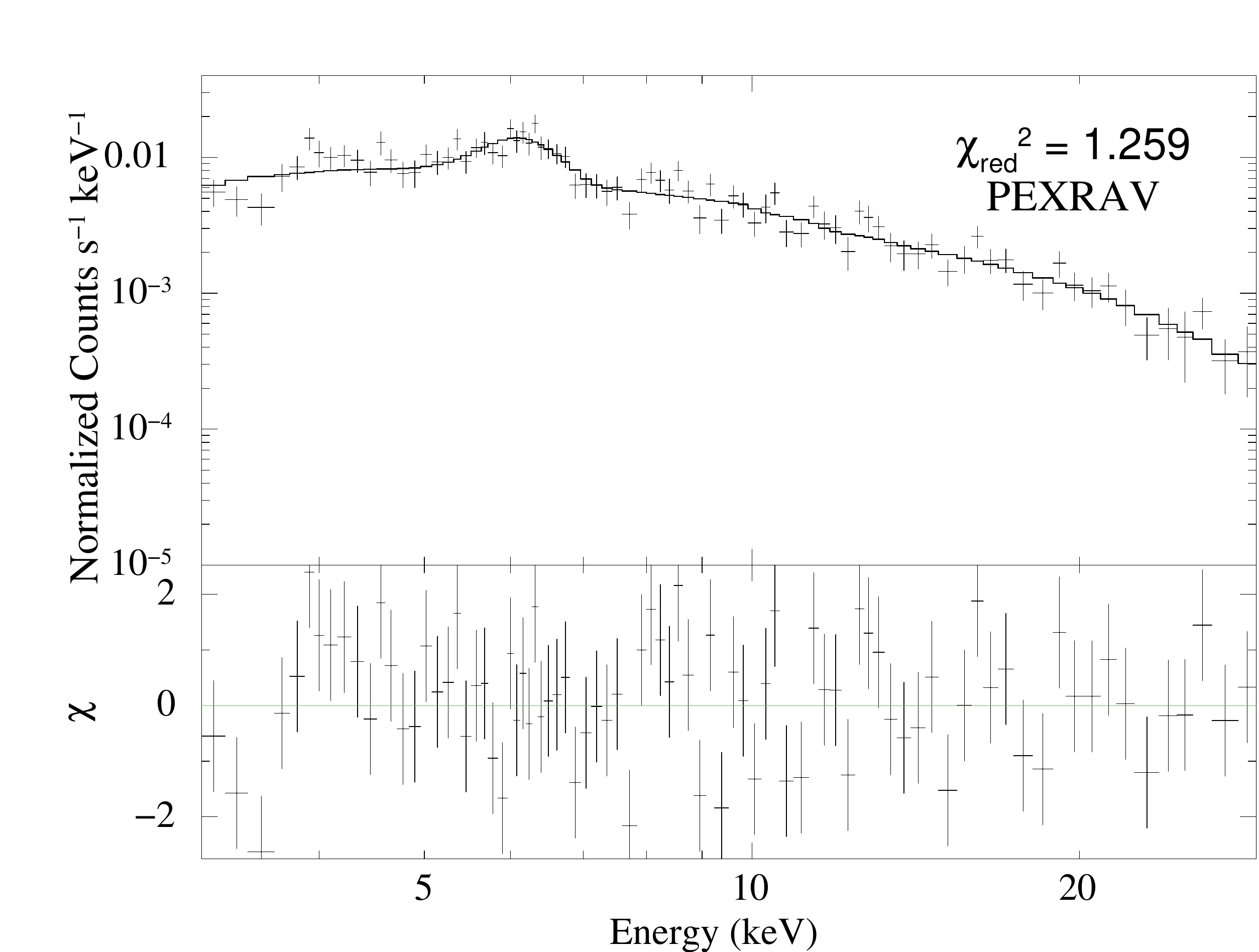}}
   \centering{
   \includegraphics[height=8truecm,angle=270]{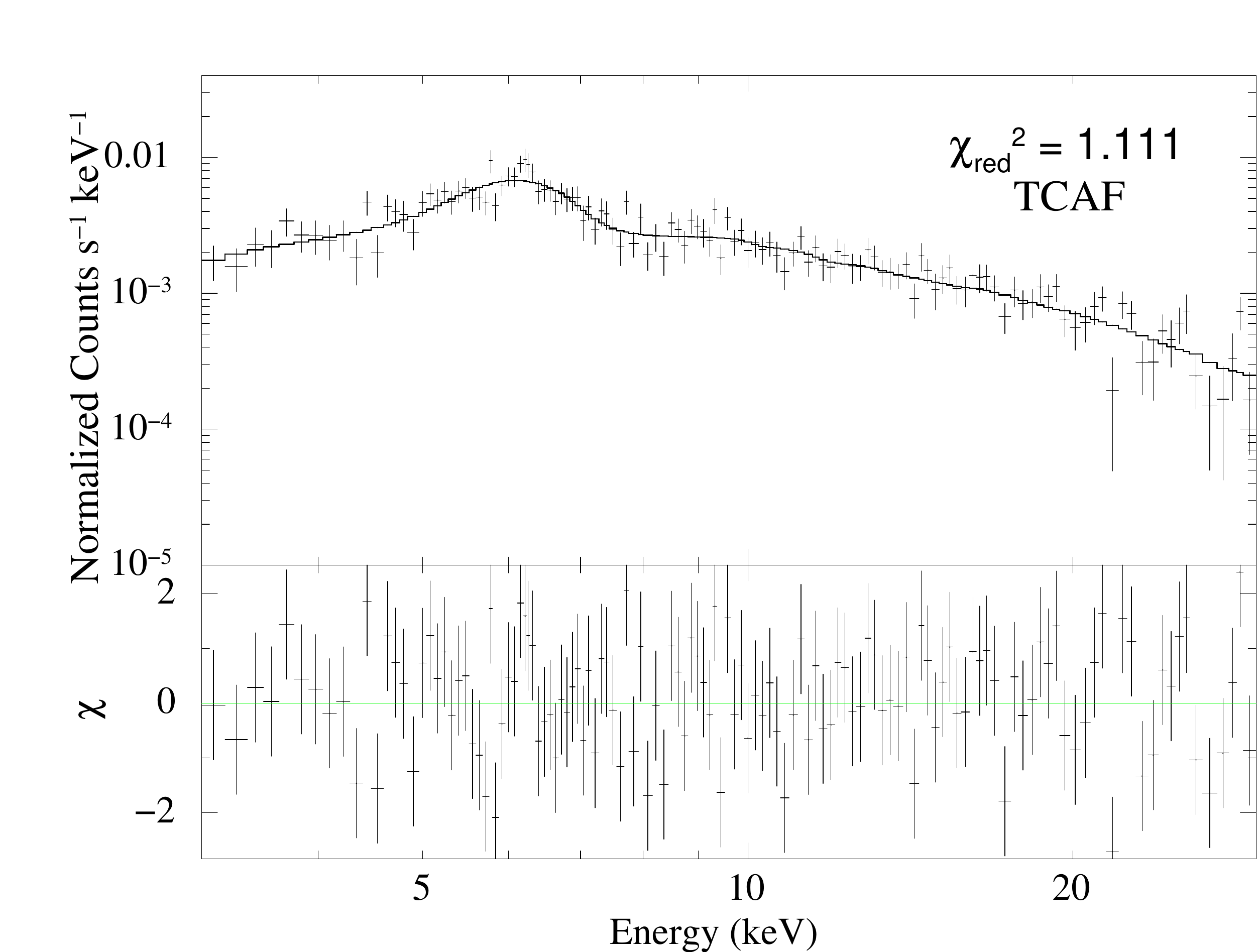}
   \includegraphics[height=8truecm,angle=270]{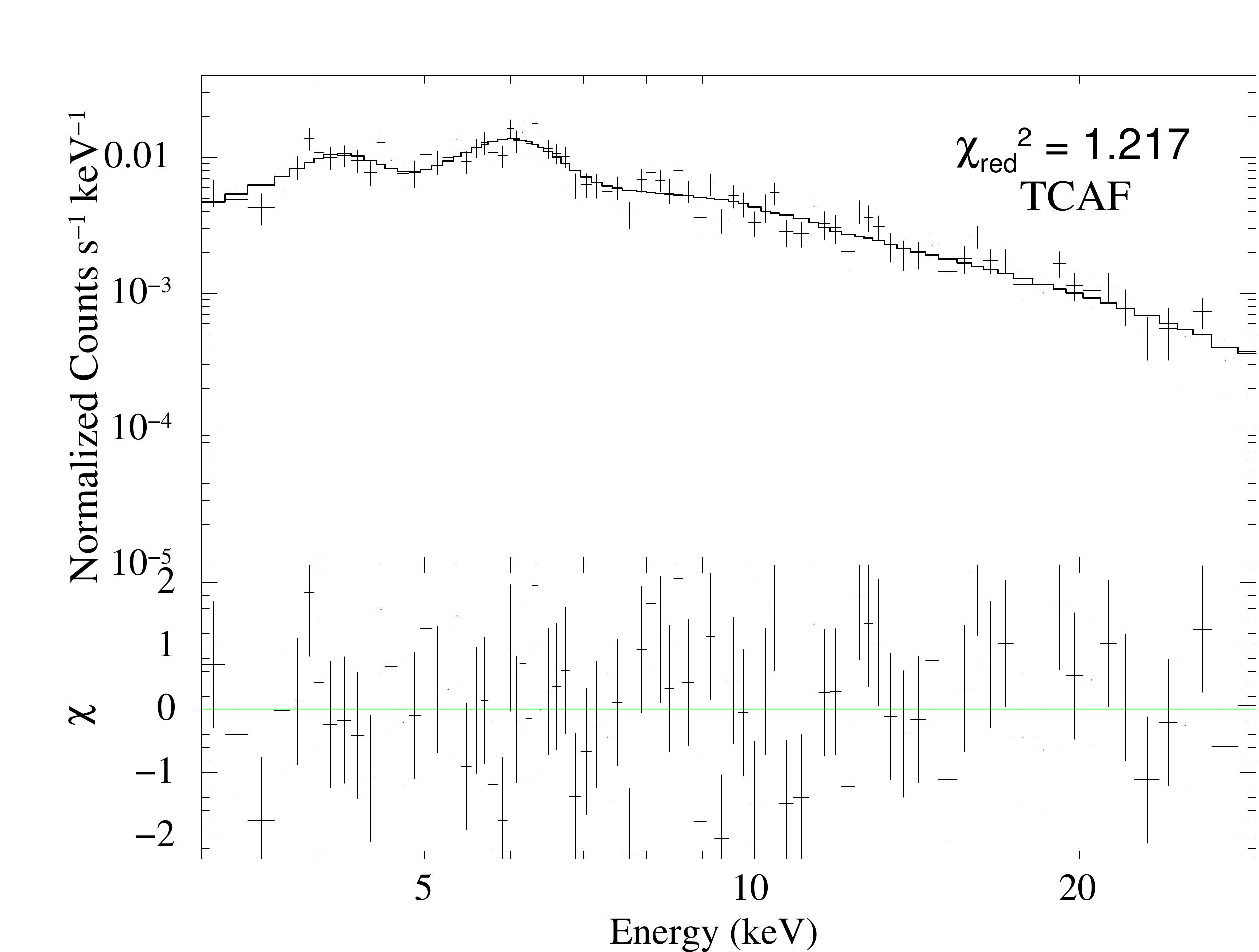}}
    \caption{Same as \autoref{fig:specPLPEXRAVTCAF} but for the OBSIDs 80201001002 (left panel) and 90602619004 (right panel) along with the residuals. \label{fig:specPLPEXRAVTCAF-3}}
\end{figure*}

\begin{figure*}
    \centering{
    \includegraphics[height=8truecm,angle=270]{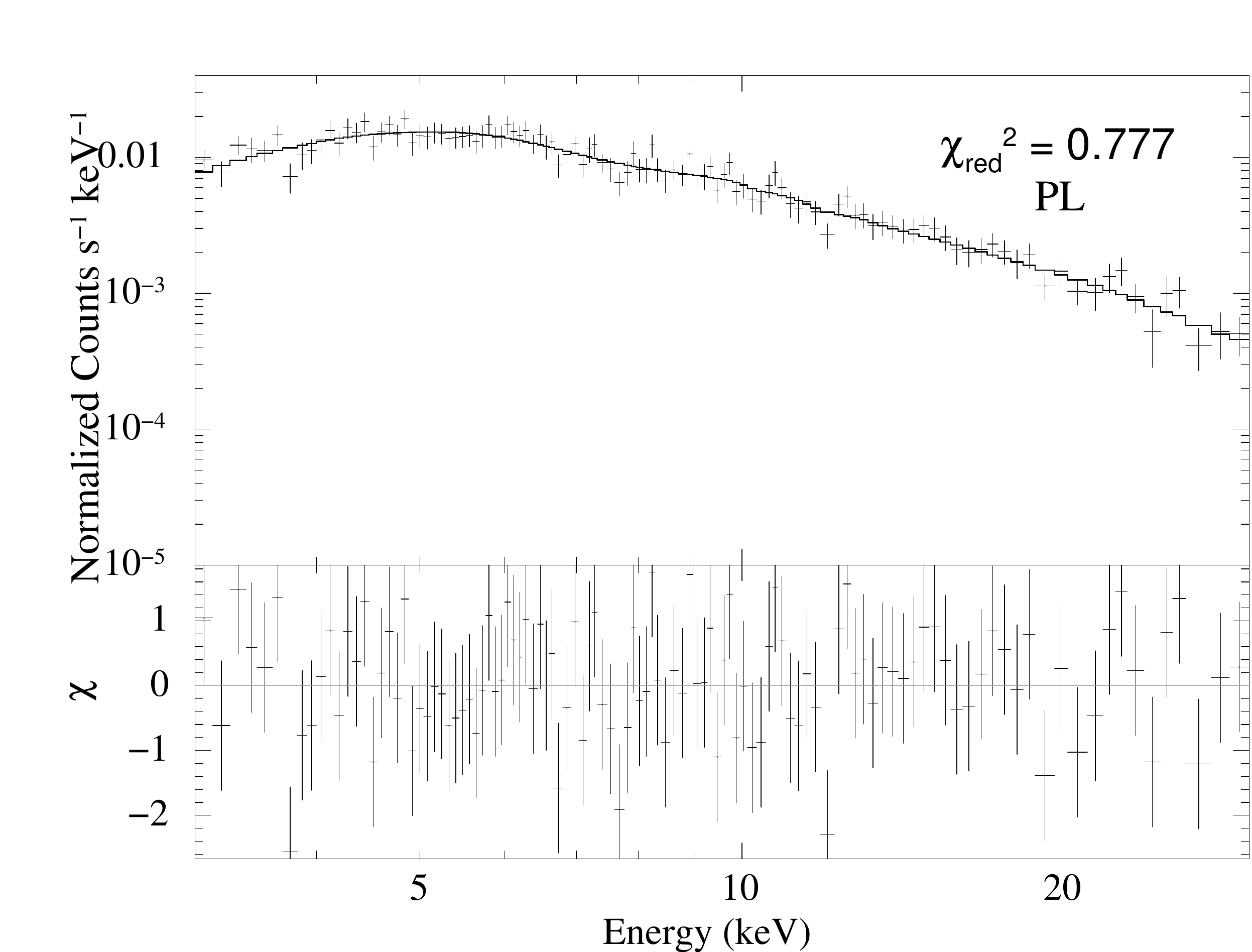}
    \includegraphics[height=8truecm,angle=270]{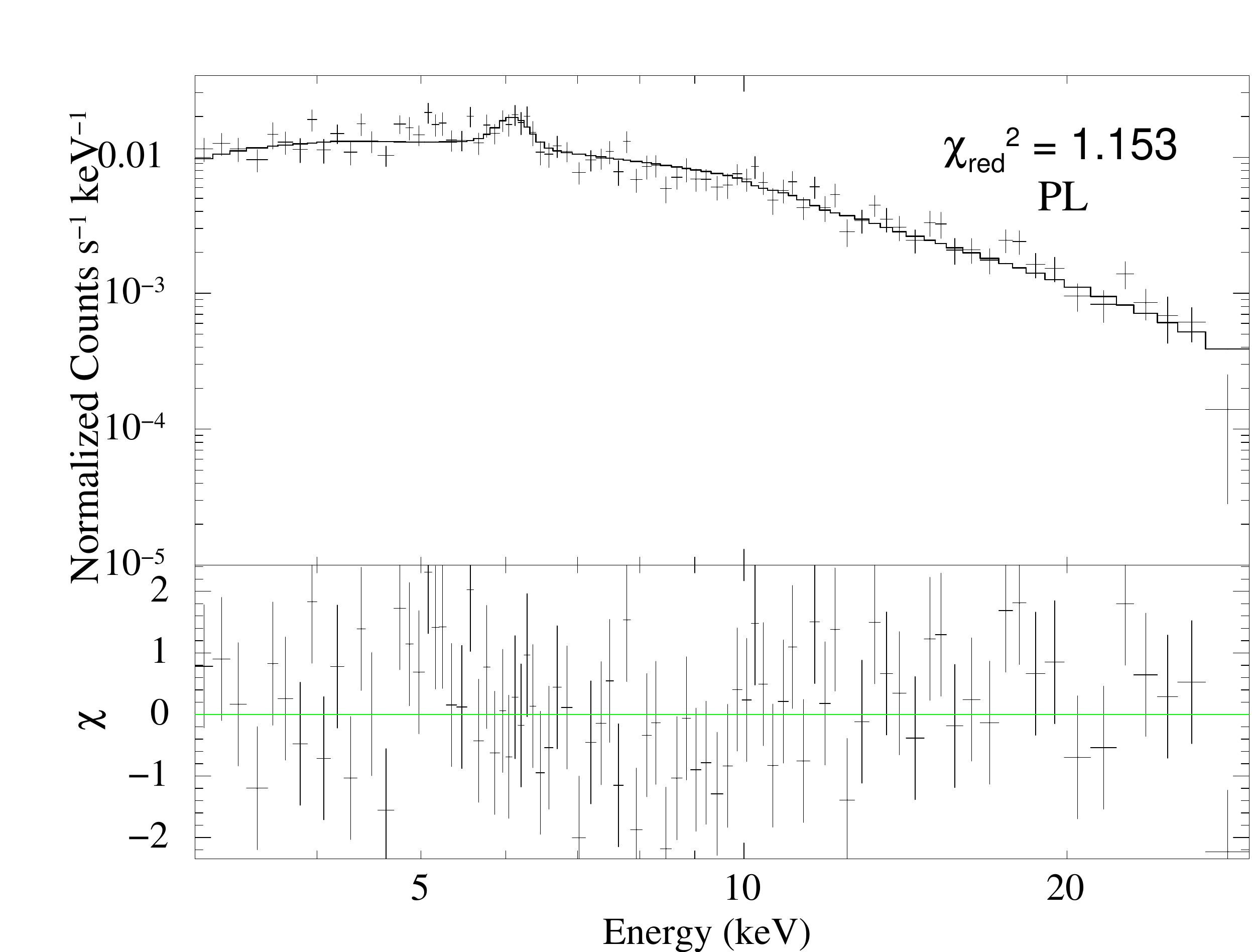}}
   \centering{
   \includegraphics[height=8truecm,angle=270]{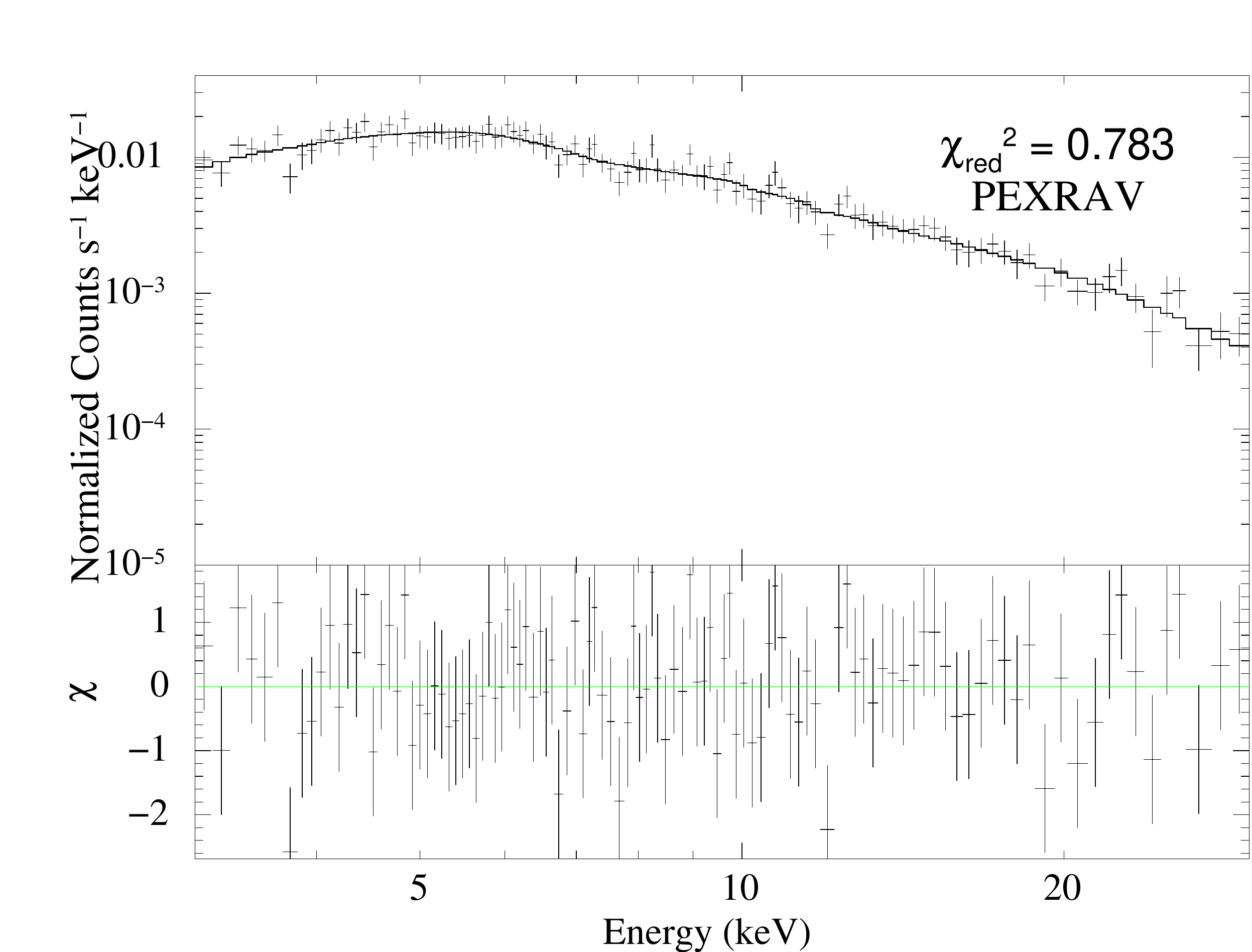}
   \includegraphics[height=8truecm,angle=270]{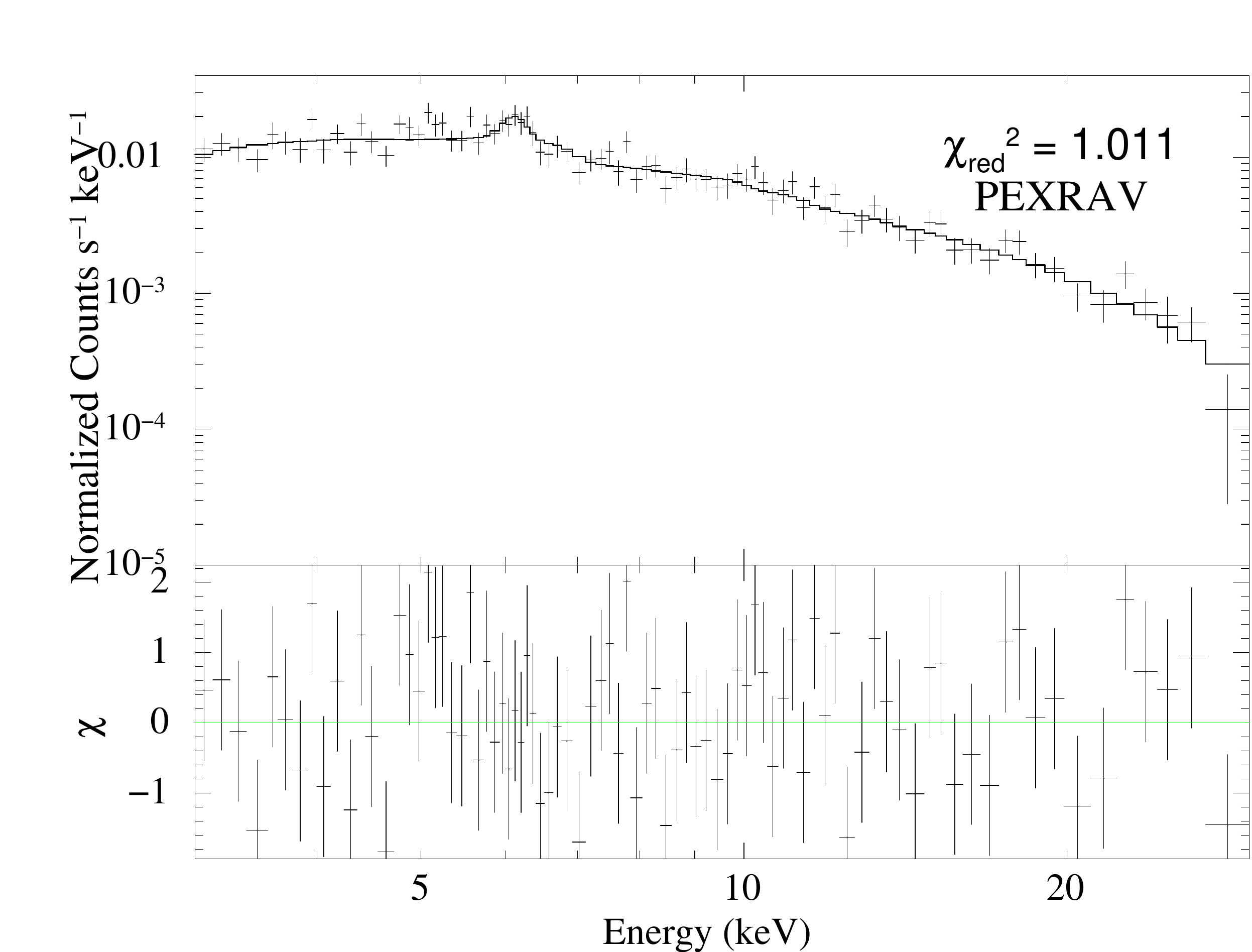}}
   \centering{
   \includegraphics[height=8truecm,angle=270]{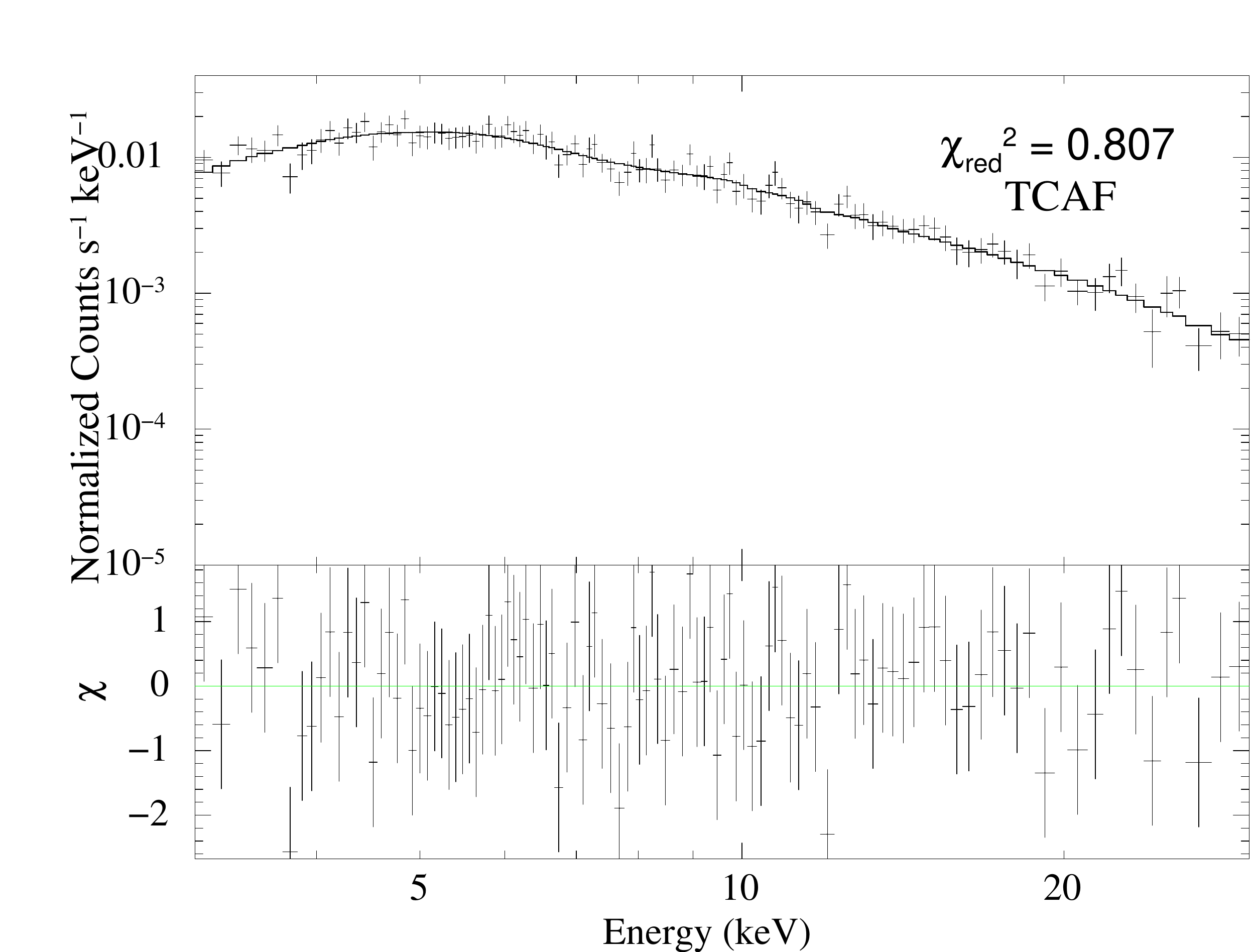}
   \includegraphics[height=8truecm,angle=270]{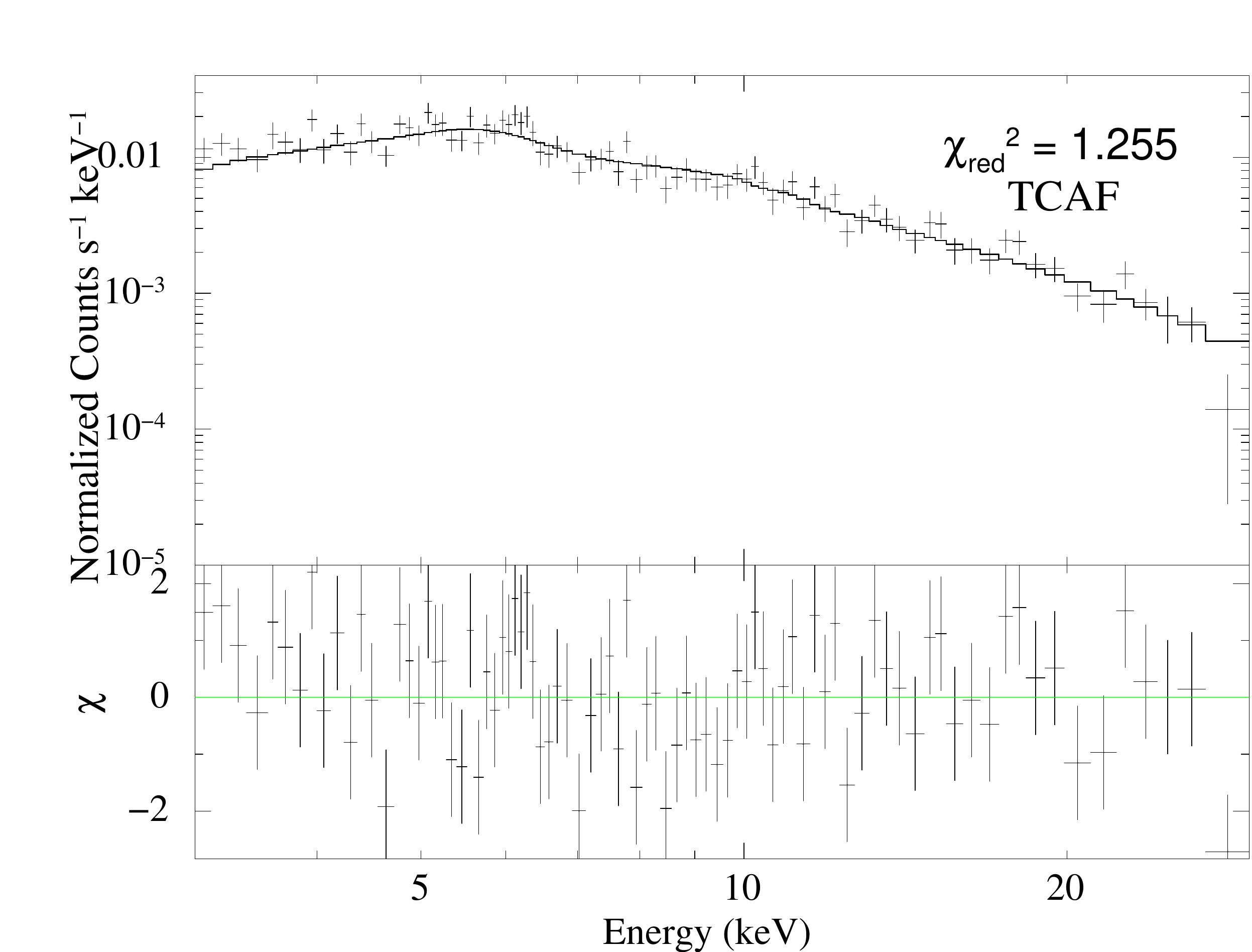}}
    \caption{Same as \autoref{fig:specPLPEXRAVTCAF} but for the OBSIDs 90602619006 (left panel) and 90602619008 (right panel) along with the residuals. \label{fig:specPLPEXRAVTCAF-4}}
\end{figure*}

\end{document}